\documentclass[
 aip,
 amsmath,amssymb,
 reprint,%
]{revtex4-1}

\usepackage{graphicx}
\usepackage{dcolumn}
\usepackage{bm}

\usepackage[utf8]{inputenc}
\usepackage[T1]{fontenc}
\usepackage{mathptmx}
\usepackage{etoolbox}
\usepackage{subcaption}
\usepackage{color}
\usepackage[whole]{bxcjkjatype}
\usepackage{siunitx}
\usepackage{comment}

\def\rewrite#1{\textcolor{black}{#1}}
\def\rewritesecond#1{\textcolor{black}{#1}}

\makeatletter
\def\@email#1#2{%
 \endgroup
 \patchcmd{\titleblock@produce}
  {\frontmatter@RRAPformat}
  {\frontmatter@RRAPformat{\produce@RRAP{*#1\href{mailto:#2}{#2}}}\frontmatter@RRAPformat}
  {}{}
}%
\makeatother
\begin{document}

\preprint{AIP/123-QED}

\title{Can plasmoid-mediated reconnection occur in collisionless systems?}
\author{Keita Akutagawa}
\email{akutagawa@eps.s.u-tokyo.ac.jp}
\affiliation{ 
Department of Earth and Planetary Science, The University of Tokyo 
7-3-1 Hongo, Bunkyo-ku, Tokyo, 113-0033, Japan
}%

\author{Shinsuke Imada}
\affiliation{ 
Department of Earth and Planetary Science, The University of Tokyo 
7-3-1 Hongo, Bunkyo-ku, Tokyo, 113-0033, Japan
}%

\author{Munehito Shoda}%
\affiliation{ 
Department of Earth and Planetary Science, The University of Tokyo 
7-3-1 Hongo, Bunkyo-ku, Tokyo, 113-0033, Japan
}%

\date{\today}

\begin{abstract}

Magnetic reconnection is a process that converts magnetic energy into plasma energy by changing the magnetic field line topology.
The outstanding question is why the reconnection rate is $\mathcal{O}(0.01 - 0.1)$ in many astrophysical phenomena, for example solar flares and terrestrial substorms.
Previous studies have shown two ideas of Hall reconnection and plasmoid instability.
However, there is no consensus on which process is the reason for the fast reconnection.
In this paper, we discuss the formation of secondary plasmoids in \rewrite{2D antiparallel collisionless reconnection} using 2.5-dimensional particle-in-cell simulations and discuss whether plasmoid-mediated reconnection occur in collisionless systems by comparing with plasmoid instability in resistive MHD simulations.
We find that in collisionless systems secondary plasmoids can indeed form.
However, the mass ratio has a strong effect on the formation of secondary plasmoids, and it indicates that secondary plasmoids do not emerge  using realistic ion-electron mass ratio ($m_i/m_e = 1836$). 
Furthermore, we find that there is no enhancement of the reconnection rate due to the secondary plasmoid in the collisionless system, as discussed in the plasmoid-mediated reconnection.
Although our simulation $\mathcal{O}(100\lambda_i)$ box is not large enough to discuss astrophysical phenomena such as solar flares, it can reflect a relatively small plasma system such as the Earth's magnetotail.

\end{abstract}

\maketitle

\section{\label{sec:level1}Introduction}

Magnetic reconnection converts magnetic energy into plasma energy through changes in magnetic field topology. It plays an important role in magnetotail substorms\cite{nagai1998, imada2007, angelopoulos2008} and solar flares\cite{tsuneta1992, imada2013, reeves2020}. Beyond the solar system, it plays a crucial role in astronomical objects such as pulsar winds and black hole jets\cite{guo2024rev}. Thus, magnetic reconnection is a universal physical process that plays a fundamental role across a wide range of scales in the universe.

A unique characteristic of magnetic reconnection is its occurrence across a wide range of scales. For example, the Earth's magnetotail has a size of approximately $10^3\lambda_i$ ($\lambda_i$ being the ion inertial length), where kinetic effects are significant throughout the system, and it behaves as a fully collisionless system. In contrast, solar flares have a size of approximately $10^{7-8} \lambda_i$, where the system as a whole follows the MHD behavior. However, the diffusion region is much smaller than the electron mean free path ($10^5\lambda_i$), making the system globally collisional but locally collisionless. Understanding magnetic reconnection requires insights across these broad scales and varying collisionality conditions. The details of scales and physical parameters in various magnetic reconnection are summarized in [H. Ji \& W. Daughton, 2011, Ref. \onlinecite{ji2011}].

One of the most important unresolved issues in magnetic reconnection is how fast magnetic reconnection can be realized on the MHD scale. In the classical Sweet-Parker model \cite{sweet1958, parker1957}, the reconnection rate, which serves as an indicator of energy conversion efficiency, scales as $S^{-1/2}$ with respect to the Lundquist number $S := \mu_0 V_A L_{CS} / \eta$, where $V_A$ is the Alfv\'en speed, $L_{CS}$ is the length of the current sheet and $\eta$ is the resistivity. However, this dependence fails to explain solar flares occurring in the solar corona ($S \sim 10^{14}$) with a reconnection rate of $\mathcal{O}(0.01-0.1)$, necessitating additional physical mechanisms for faster reconnection.

A classical solution to fast reconnection (with a reconnection rate of $\mathcal{O}(0.01 - 0.1)$) is the Petschek model \cite{petschek1965}, where slow-mode shocks facilitate efficient magnetic energy diffusion. However, its implementation requires localized anomalous resistivity\cite{kulsrud2011}, the physical origin of which remains unclear. A model that explains fast reconnection without anomolous resistivity is the plasmoid-mediated reconnection model. In systems with a sufficiently large Lundquist number ($S > S_c = \mathcal{O}(10^4)$), current sheets induce plasmoid instability\cite{loureiro2007, bhattacharjee2009, shibayama2015, zenitani2020}. This instability results in the creation and growth of plasmoids over various scales, leading to fast magnetic reconnection with a rate of about 0.01\cite{shibata2001, uzdensky2010, ji2011}. The Lundquist number of the solar corona ($S \sim 10^{14}$) is well above the threshold for plasmoid instability, making plasmoid-mediated reconnection a significant process for explaining solar flares. Indeed, signatures of plasmoid instability are sometimes observed during flare events \cite{takasao2012}.
 
In the kinetic regime, in 2001 the GEM Challenge \cite{birn2001} revealed that kinetic effects (including Hall effect) should be important for fast magnetic reconnection. They carried out four different simulations, which are Resistive MHD with uniform resistivity, Hall MHD, Hybrid (particle ions and fluid electrons), PIC (particle ions and electrons) and showed that the reconnection rate was $\mathcal{O}(0.1)$ when using Hall MHD, Hybrid, PIC. On the other hand, the reconnection rate in Resistive MHD is much smaller than others. The major differences between Hall reconnection (seen in models which include Hall effect) and Sweet-Parker reconnection (seen in Resistive MHD with uniform resistivity) appear in the following points. First, quadrupole magnetic field is seen\cite{nagai2001, pritchett2001}. It is perpendicular to the plane of anti-parallel magnetic field. Second, the decoupling of the motion of ions and electrons around the diffusion region is seen and it creates current\cite{pritchett2001, treumann2013}. Quadrupole magnetic field corresponds to this current\cite{drake2009}. Third, the decoupling of the distribution of ions and electrons is seen and it creates electric field because of the charge separation\cite{zenitani2011pop}. This electric field corresponds to the Hall term in the Ohm's law. Finally, the bifurcated current sheet is seen in the outflow region\cite{hoshino1998}. The detail of these structure is written in section \ref{chap2}.

In systems where the diffusion region becomes collisionless, such as in solar flares, \rewrite{dissipation} is determined by kinetic effects, leading to a behavior distinct from the MHD description. Consequently, it is not obvious whether fast reconnection, like Petschek or plasmoid-mediated reconnection, occurs within the collisonless regime. In addition, when kinetic effects are important, fast magnetic reconnection (with the reconnection rate of $\sim \mathcal{O}(0.1)$) can be achieved by the Hall effect\cite{birn2001} (in the absence of slow shocks or plasmoids), though the reason why the Hall effect accelerates reconnection is not immediately apparent from previous studies (electron-positron system\cite{liu2018eleposi}, strong guide field system\cite{liu2014guide}). Recently, the pressure balance around the diffusion region has been found to be important\cite{liu2017why, liu2022nat}. It is important to investigate whether magnetic reconnection, with a diffusion region governed by kinetic scales, exhibits the same macroscopic behavior as the MHD model.

Some full particle simulations reported that secondary plasmoids were formed in the reconnecting current sheet in collisionless plasma. 
\rewrite{Especially in PIC simulations for pair plasmas, it is known that secondary plasmoids are well formed\cite{cerutti2013, sironi2016}. From these results, mass ratio $m_i / m_e$ should be important for secondary plasmoid formation\cite{daughton2006}, however the reason is not yet clear.}
It is also known that guide field promotes the production of secondary plasmoid by Kelvin-Helmholtz instability\cite{drake2006grl, fermo2012}. The PIC simulation results that generate these secondary plasmoids are very similar to the plasmoid instability. On the other hand, no studies discuss whether the secondary plasmoid formed in collisionless magnetic reconnection is physically equivalent to the plasmoid reconnection discussed in Resistive MHD framework or not.

{This study aims to examine whether plasmoid-mediated reconnection, described in the framework of MHD, occurs in collisionless plasmas, where resistivity is governed by kinetic effects. Previous PIC simulations have reported the formation of structures like secondary plasmoids, a characteristic of plasmoid instability, during collisionless anti-parallel magnetic reconnection\cite{daughton2006, oka2008}. However, it remains unclear whether these structures exhibit the same physical properties as plasmoid-mediated reconnection. In this study, we employ 2.5-dimensional PIC simulations to explore the parameter dependence of secondary plasmoids, and compare the results with MHD simulations to assess whether the secondary plasmoids observed in PIC simulations are physically equivalent to those in MHD simulations.

\section{\label{chap2}Simulation Setup}

The simulation code used in this study is the open-source plasma simulation \rewrite{code} 'KAMMUY' developed by us. It includes both PIC and MHD simulation codes, written in CUDA and C++ for GPU acceleration.

The PIC code employs Yee-lattice, Leapfrog integration, and Langdon-Marder type correction\cite{marder1987, langdon1992} to preserve charge conservation. The MHD code is based on the finite volume method with the HLLD Riemann solver\cite{miyoshi2005}, 2nd-order MUSCL reconstruction\cite{vanleer1979}, and a 2nd-order Runge-Kutta method. The solenoidal condition of the magnetic field is enforced using the upwind constrained transport method\cite{gardiner2005}.

\subsection{\label{chap2:setup}PIC simulation setup}

We use Harris current sheet\cite{harris1962} with an initial magnetic field $B_x = B_0 \tanh (y / \delta), B_y = 0, B_z = 0$ (no guide field) and a number density $n = n_0 \cosh^{-2}(y/\delta) + n_b$; here $\delta $ is the half-thickness of the current sheet and $n_b$ is the number density of background plasmas. The temperature ratio on the current sheet is $T_i / T_e = 1$, and the ratio of the electron plasma frequency to the electron cyclotron frequency is $\omega_{pe} / \Omega_{ce} = 1$. The half-thickness is set to $\delta = 2.0\lambda_i$. The number density and temperature of the background plasmas are $n_b = 0.2 n_0$ and $T_{ib} = T_{eb} = 0.2 T_i$ so that the \rewrite{upstream} plasma beta is $\beta = 0.04$. We use 100 ppc on the current sheet (20 ppc at background region)

The simulation box size is fixed at $L_x \times L_y = 100\lambda_i \times 50\lambda_i $ where $\lambda_i$ is the ion inertial length. To verify that our results are not affected by boundary effects, we also performed simulations with $L_x \times L_y = 200 \lambda_i \times 100 \lambda_i $. The boundaries are periodic in the x-direction and conducting walls in the y-direction. The grid spacing and time step are set to $\Delta x = \Delta y = 0.1\lambda_e$ (where $\lambda_e$ denotes the electron inertial length) and $\Delta t = 0.05 \omega_{pe}^{-1}$, respectively. Reconnection is initiated by a small perturbation in the z-component of the vector potential, given by $\Delta A_z = -2 \delta \exp (-(x^2 + y^2) / (2\delta)^2) \times \epsilon B_0$, where $\epsilon$ is set to 0.1. 

The evolution of reconnection is examined over a range of mass ratios ($m_i/m_e = 1-32$). A summary of the simulation setup is provided in Table \ref{chap2:setup_table}.

\begin{table}[htbp]
\caption{\label{chap2:setup_table}PIC simulation setup}
\begin{ruledtabular}
\begin{tabular}{ccccc}
Run & $m_i / m_e$ & Domain size & Grid cells & Particles \\
\hline
  1 & 1  & $100\lambda_i \times 50\lambda_i$  & $1000 \times 500$  & $2.8 \times 10^7$ \\
  2 & 2  & $100\lambda_i \times 50\lambda_i$  & $1414 \times 717$  & $5.6 \times 10^7$ \\
 3A & 4  & $100\lambda_i \times 50\lambda_i$  & $2000 \times 1000$ & $1.1 \times 10^8$ \\
 3B & 4  & $200\lambda_i \times 100\lambda_i$ & $4000 \times 2000$ & $3.8 \times 10^8$ \\
  4 & 8  & $100\lambda_i \times 50\lambda_i$  & $2828 \times 1414$ & $2.2 \times 10^8$ \\
 5A & 16 & $100\lambda_i \times 50\lambda_i$  & $4000 \times 2000$ & $4.5 \times 10^8$ \\
 5B & 16 & $200\lambda_i \times 100\lambda_i$ & $8000 \times 4000$ & $1.5 \times 10^9$ \\
  6 & 32 & $100\lambda_i \times 50\lambda_i$  & $5656 \times 2828$ & $9.0 \times 10^8$ \\
\end{tabular}
\end{ruledtabular}
\end{table}

\subsection{\label{chap3:setup}Resisitive MHD simulation setup}

We use Harris current sheet\cite{harris1962} with an initial magnetic field $B_x = B_0 \tanh (y / \delta), B_y = 0, B_z = 0$ (no guide field) and a density $\rho = \rho_0 \cosh^{-2}(y/\delta) + \rho_b$; here $\delta$ is the half-thickness of the current sheet and $\rho_b$ is the density of background plasma. The half-thickness is set to $\delta = 1.0$. The density and temperature of the background plasmas are $\rho_b = 0.5 \rho_0$ and $T_b = 0.5 T$ so that the \rewrite{upstream} plasma beta is $\beta = 0.25$.

The size of the simulation box is $L_x \times L_y = 400 \times 80$ and the resistivity is set to $\eta = 10^{-3}$ so that the Lundquist number is greater than the threshold of plasmoid instability $S_{CS} \sim \mathcal{O}(10^4)$. To verify that our results are not affected by boundary effects, we also performed simulations with $L_x \times L_y = 1000 \times 200$. The size of the grid is $\Delta x = \Delta y = 1.0/16.0$. Initial current sheet is pressed by small incompressible perturbation; Reconnection is triggered by a small perturbation; $d A_z = -2 \delta \exp (-(x^2 + y^2) / (2\delta)^2) \times \epsilon B_0$ and $\epsilon$ is set to 0.05. 

A summary of our simulation setup is given in Table \ref{chap3:setup_table}.

\begin{table}[htbp]
\caption{\label{chap3:setup_table}Resistive MHD simulation setup}
\begin{ruledtabular}
\begin{tabular}{ccccc}
Run & Domain size & Grid size & Grid cells\\
\hline
  7 & $400 \times 80$ &  $1.0 / 16.0$ & $6400 \times 1280$ \\
  8 & $1000 \times 200$ &  $1.0 / 16.0$ & $16000 \times 3200$ \\
\end{tabular}
\end{ruledtabular}
\end{table}

\section{\label{chap3}Secondary plasmoid in collisionless magnetic reconnection}


\subsection{\label{chap3:overview}Overview}

\begin{figure}[htbp]
    \includegraphics[width=\linewidth]{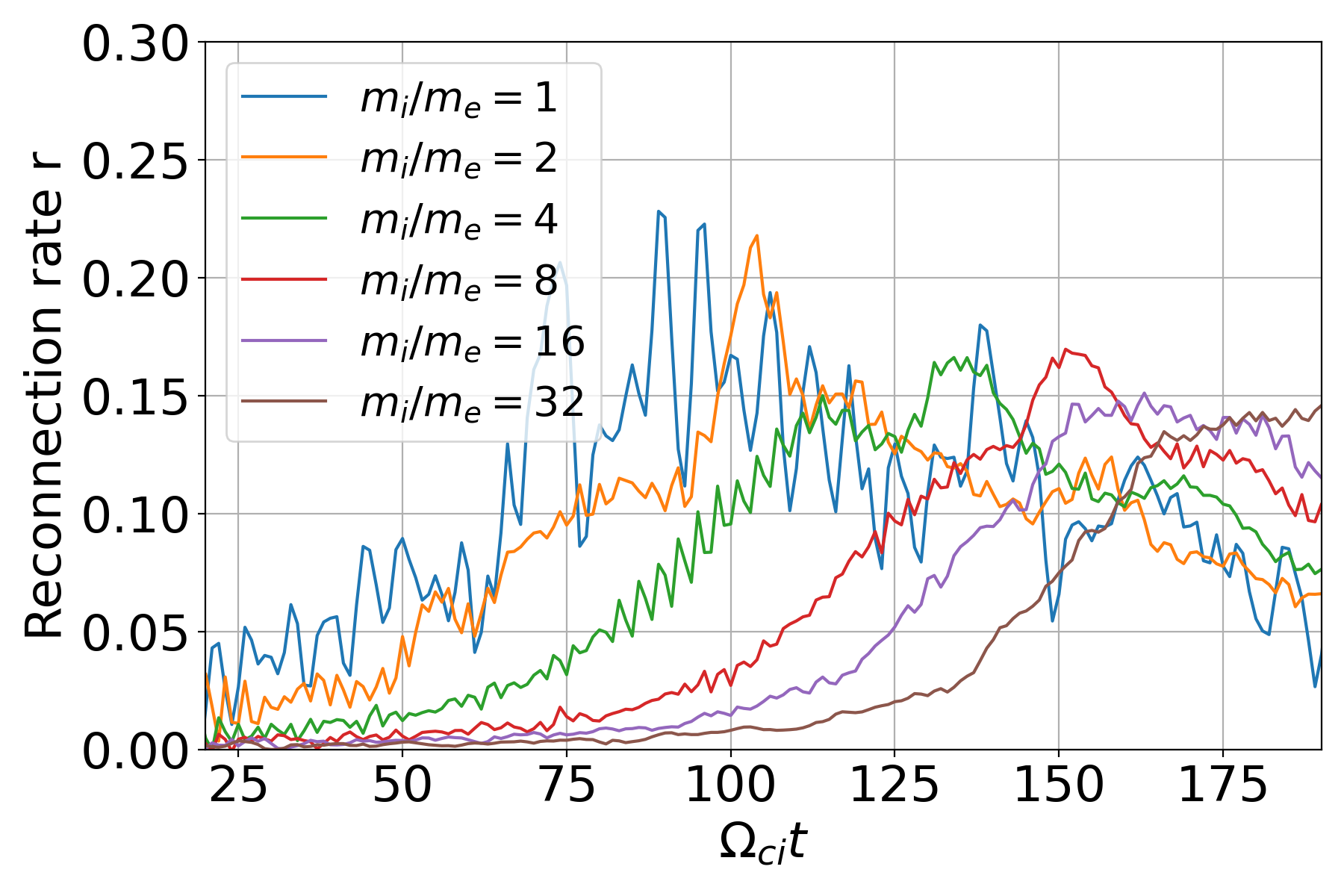}
    \caption{\label{chap3:reconnection_rate_fig}Time evolution of reconnection rate calculated by \eqref{chap3:reconnection_rate_eq} for each $m_i / m_e$ case.}
\end{figure}

\begin{figure}[htbp]
    \includegraphics[width=\linewidth]{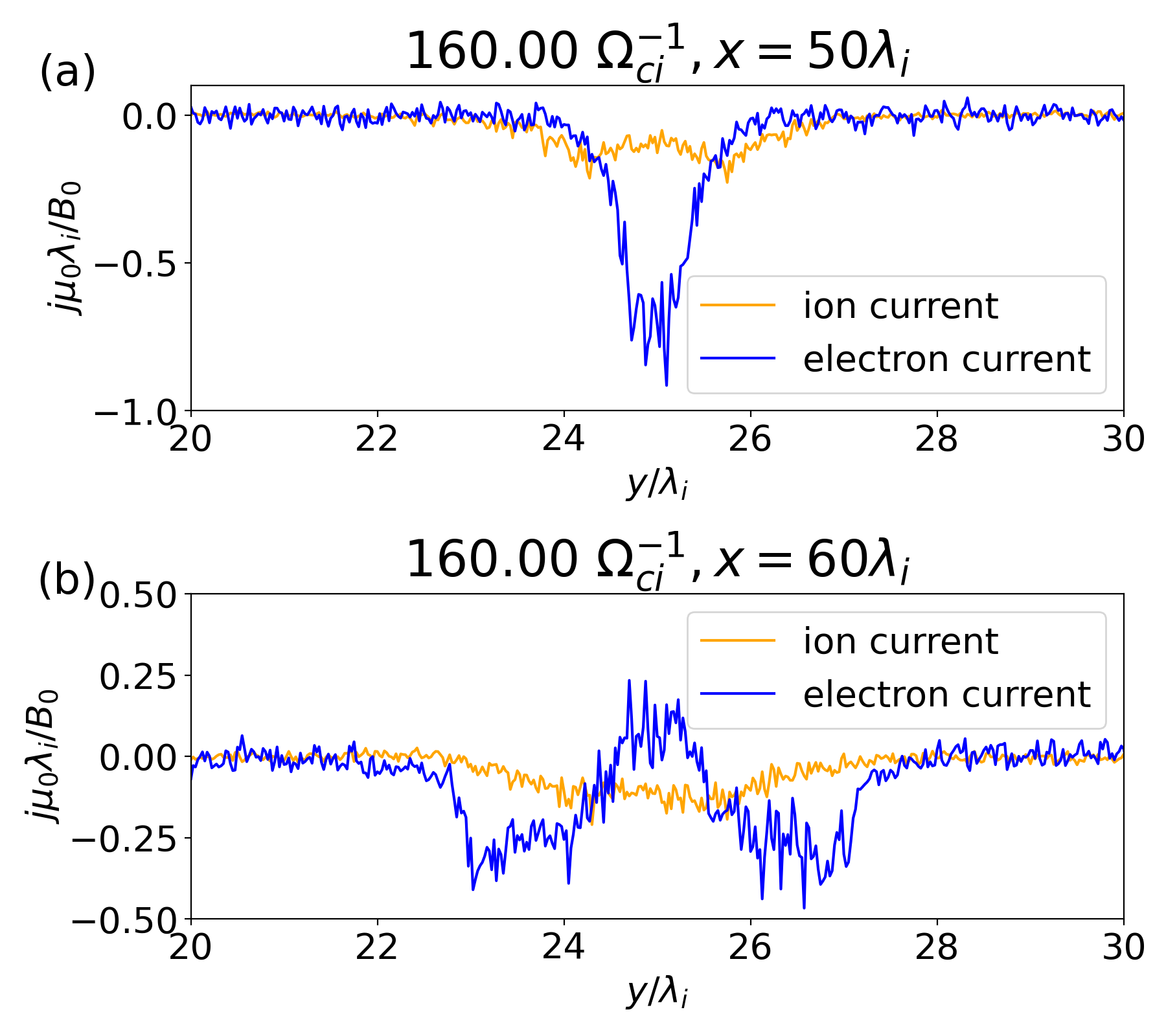}
    \caption{\label{chap3:current_contribution_fig}Ions and electrons contribution to the out-of-plane current $J_z$ at $x = 50 \lambda_i, 60\lambda_i$ for $m_i / m_e = 16$ case (Run 5A) at $160 \Omega_{ci}^{-1}$.}
\end{figure}

\begin{figure*}[htbp]
    \includegraphics[width=\textwidth]{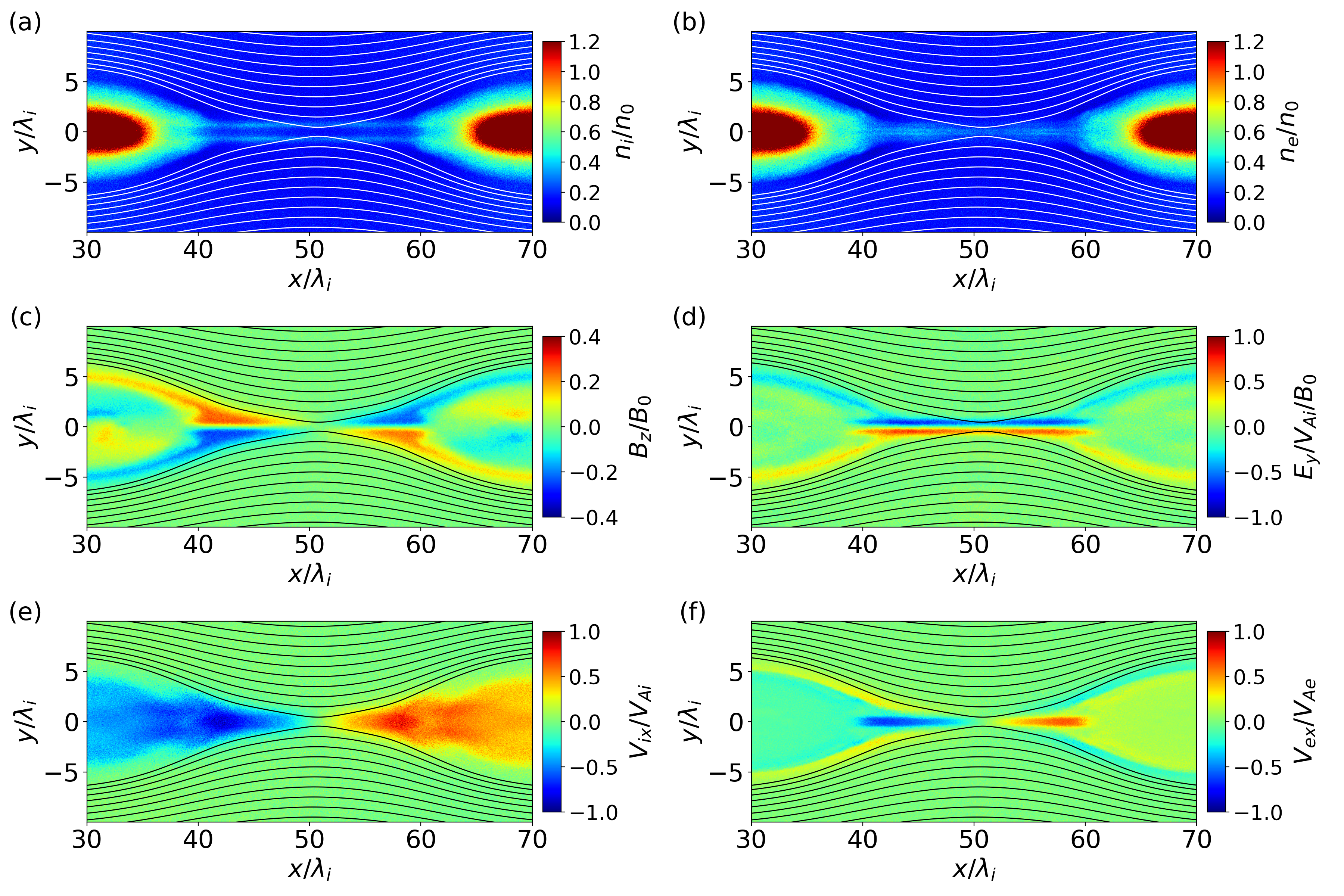}
    \caption{\label{chap3:overview_fig}Spatial distributions of physical variables for $m_i / m_e = 16$ case (Run 5A) at steady state ($t = 160 \Omega_{ci}^{-1}$). (a) number density of ions $n_i$ normalized by initial number density $n_0$, (b) number density of electrons $n_e$ normalized by $n_0$, (c) out-of-plane magnetic field $B_z$ normalized by initial anti-parallel magnetic field $B_0$, (d) the Hall electric field $E_y$ normalized by $V_{Ai} B_0$, where $V_{Ai} := B_0 / \sqrt{\mu_0 n_0 m_i}$, (e) ion bulk speed in the x direction $V_{ix}$ normalized by $V_{Ai}$, (f) electron bulk speed in the x direction $V_{ex}$ normalized by $V_{Ae} := B_0 / \sqrt{\mu_0 n_0 m_e}$. White or black lines show the in-plane magnetic field lines.}
\end{figure*}

Figure \ref{chap3:reconnection_rate_fig} presents the reconnection rate for each $m_i / m_e$ cases. The reconnection region is identified as the minimum point of the reconnecting magnetic flux, defined as $\Phi_{\rm min}(t) := \min (\Phi(x, t))$, where $\Phi(x, t) := \int |B_x(x, y, t)| dy$. The reconnection rate is calculated as
\begin{eqnarray}
r = -\frac{1}{V_{A0} B_{0}} \frac{d}{dt}\Phi_{\rm min},
\label{chap3:reconnection_rate_eq}
\end{eqnarray}
where $V_{A0} := B_0 / \sqrt{\mu_0 n_0 (m_i + m_e)}$ is the Alfv\'en speed, and $B_0$ is the anti-parallel magnetic field strength. Reconnection starts at $t = 100 \Omega_{ci}^{-1}$ and reaches a steady rate of about 0.15 after $t = 160 \Omega_{ci}^{-1}$. This value closely matches that found in previous studies \cite{birn2001, zenitani2011pop, liu2018eleposi}. The variation in the reconnection onset time is attributed to ion mass: lower $m_i / m_e$ allows more ions to contribute to dissipation through Landau resonance with the out-of-plane electric field $E_z$. The delayed transition to steady state compared to previous studies is attributed to the use of a relatively thicker initial current sheet ($\delta = 2\lambda_i$).

Figure \ref{chap3:current_contribution_fig} shows the contributions of ions and electrons to the out-of-plane current $j_z$ at different positions. Figure \ref{chap3:current_contribution_fig}(a) presents the contributions at $x = 50\lambda_i$, where electrons mainly contribute to $j_z$. The ion contribution exhibits two peaks corresponding to the two peaks in number density (see Figure \ref{chap3:overview_fig}(a)), but its magnitude is small and has little impact on the total $j_z$. Figure \ref{chap3:current_contribution_fig}(b) illustrates the contributions at $x = 60\lambda_i$, near the edge of the Hall effect and outflow jet (see Figure \ref{chap3:overview_fig}). The electron contribution exhibits two peaks, resulting in a bifurcated current sheet, which has been observed in both simulations and observations\cite{hoshino1998, runov2003}.

Figure \ref{chap3:overview_fig} shows the spatial distributions of physical variables for the $m_i / m_e = 16$ case (Run 5A) at $t = 160 \Omega_{ci}^{-1}$. Figure \ref{chap3:overview_fig}(a) and (b) display the number densities of ions and electrons. The ion density exhibits two peaks near the center of the reconnection region ($x \sim 50\lambda_i, y \sim 0\lambda_i$), whereas the electron density has only one peak. The difference in spatial distribution between ions and electrons induces charge separation, as reported in a previous study\cite{zenitani2011pop}. Figure \ref{chap3:overview_fig}(c) presents the out-of-plane magnetic field $B_z$, referred to as the Hall magnetic field, which has a quadrupole structure. Its magnitude is approximately $0.1 B_0$, with a sign reversal in the outflow regions ($x = [30\lambda_i, 40\lambda_i], [60\lambda_i, 70\lambda_i]$). Both the magnitude and sign are consistent with previous studies\cite{drake2009}. Figure \ref{chap3:overview_fig}(d) presents the electric field $E_y$, known as the Hall electric field, whose sign reverses at $y \sim 0$. This field corresponds to the charge separation shown in Figure \ref{chap3:overview_fig}(a) and (b). Its magnitude ranges in $0.5 - 1.0 V_{Ai} B_0$. A previous study\cite{zenitani2011pop} reported a value of $\sim V_{Ai} B_0$, which is consistent with our result. Figure \ref{chap3:overview_fig}(e) and (f) show the x-component of the bulk speed of ions and electrons, $V_{ix}$ and $V_{ex}$. In the outflow region, $V_{ix}$ and $V_{ex}$ are on the order of $V_{Ai}$ and $V_{Ae}$, respectively. At the boundary between the inflow and outflow regions, $V_{ex}$ exhibits a nonzero value along the magnetic field lines. While the structure of $V_{ex}$ in the outflow region resembles that of $V_{ix}$, the jet thickness differs. The spatial distributions of $V_{ix}$ and $V_{ex}$ are consistent with previous studies\cite{pritchett2001, treumann2013}.

To summarize, we find that our simulation code, originally developed for this study, can reproduce various aspects of collisionless reconnection reported in the literature (at least when $m_i/m_e=16$ is adopted). This result confirms the validity of our code for magnetic reconnection simulations and ensures the reliability of the results presented later.


\subsection{\label{chap3:mass_ratio_dependence_of_secondary_plasmoid}Mass ratio dependence of secondary plasmoids}

\begin{figure*}
    \includegraphics[width=0.95\textwidth]{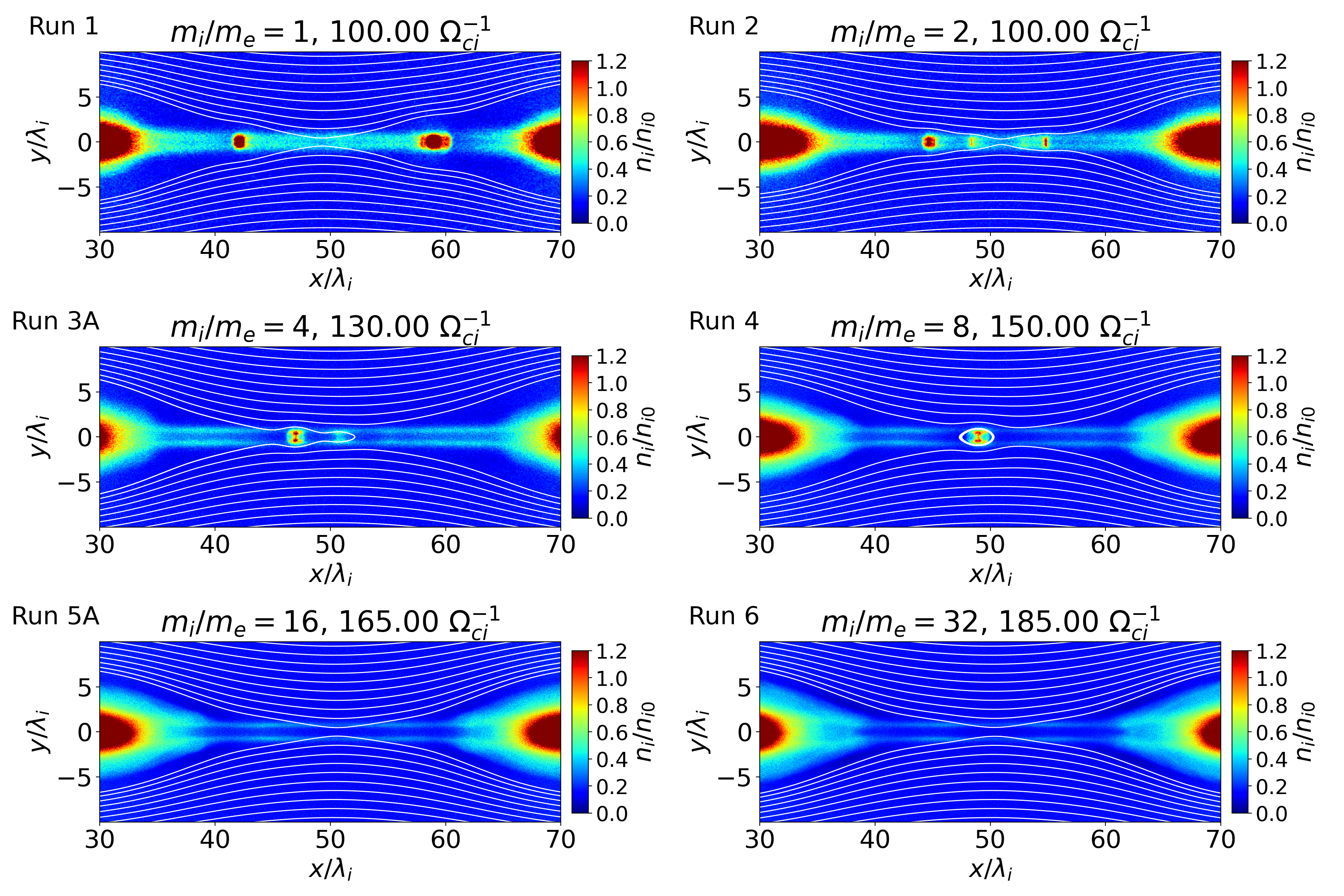}
    \caption{\label{chap3:whole_results}Spatial distribution of the number density of ions for each $m_i / m_e$ case normalized by the initial value $n_{i0}$. We choose snapshots which has secondary plasmoids. For $m_i / m_e = 16, 32$ cases, we cannot find any secondary plasmoids in whole simulation time. White line shows the magnetic field lines.}
\end{figure*}

Figure \ref{chap3:whole_results} presents the ion number density for various $m_i / m_e$ cases. Secondary plasmoids are distinctly observed for $m_i / m_e = 1, 2, 4, 8$ cases, whereas they are not observed for $m_i / m_e = 16, 32$ cases. Additionally, the number of secondary plasmoids increases as $m_i / m_e$ decreases. These findings suggest that secondary plasmoid formation depends on $m_i / m_e$, with a possible threshold between $m_i / m_e = 8$ and $m_i / m_e = 16$.
    
We found that in collisionless plasma systems secondary plasmoids could indeed form. However, the $m_i / m_e$ has a strong effect on the formation of secondary plasmoids, and it is difficult to form secondary plasmoids using high $m_i / m_e$.

\subsection{\label{chap3:collisionless_tearing_mode_in_electron_diffusion_region}Collisionless tearing mode in electron diffusion region}

\begin{figure}[htbp]
    \includegraphics[width=0.95\linewidth]{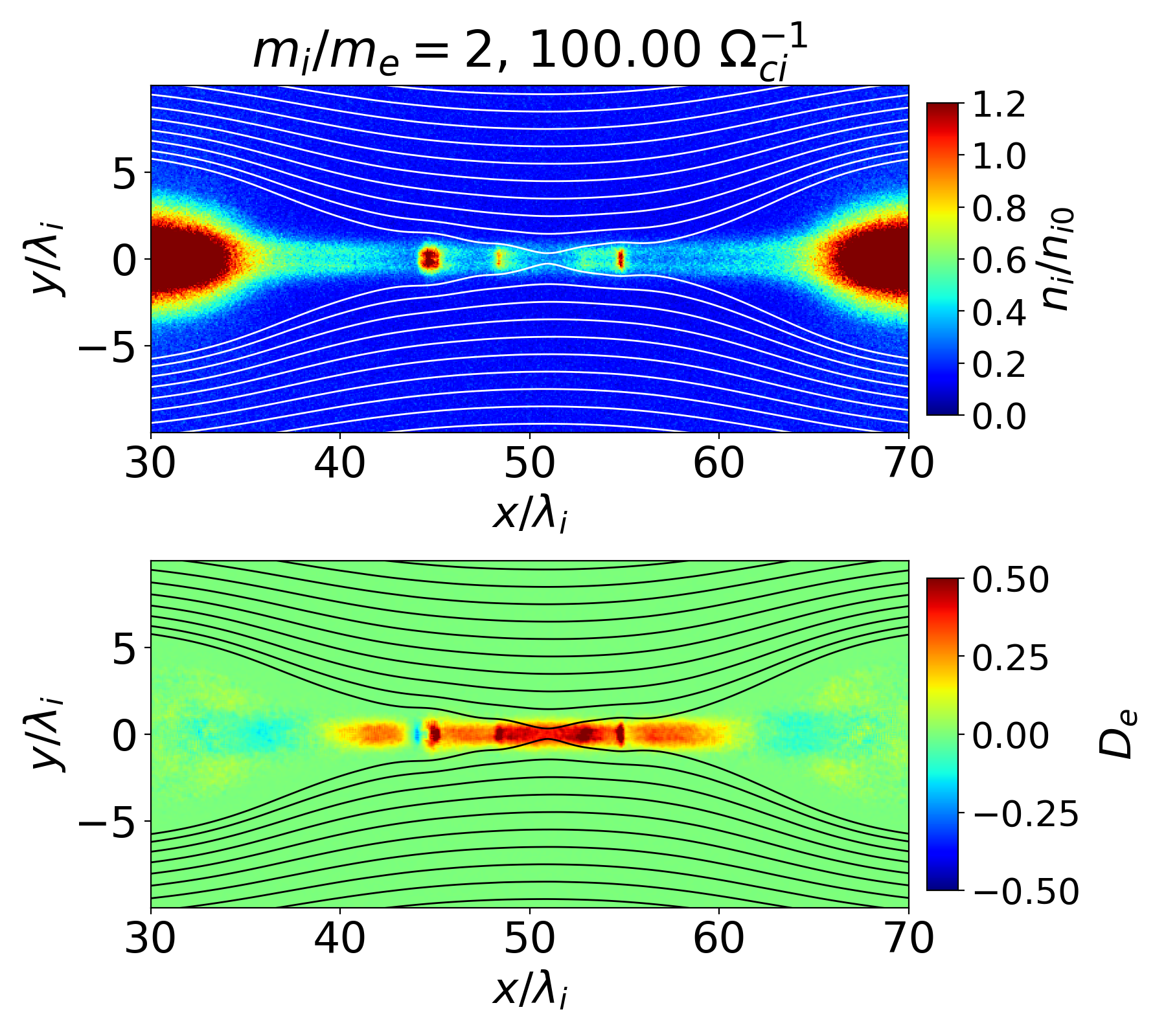}
    \caption{\label{chap3:de_with_plasmoid}Enlarged spatial distributions of the ion number density and $D_e$ at electron rest frame for $m_i / m_e = 2$ case. Several secondary plasmoids are found inside the electron diffusion region ($D_e > 0$). White or black lines show the in-plane magnetic field lines.}
\end{figure}

\begin{figure}[htbp]
    \includegraphics[width=\linewidth]{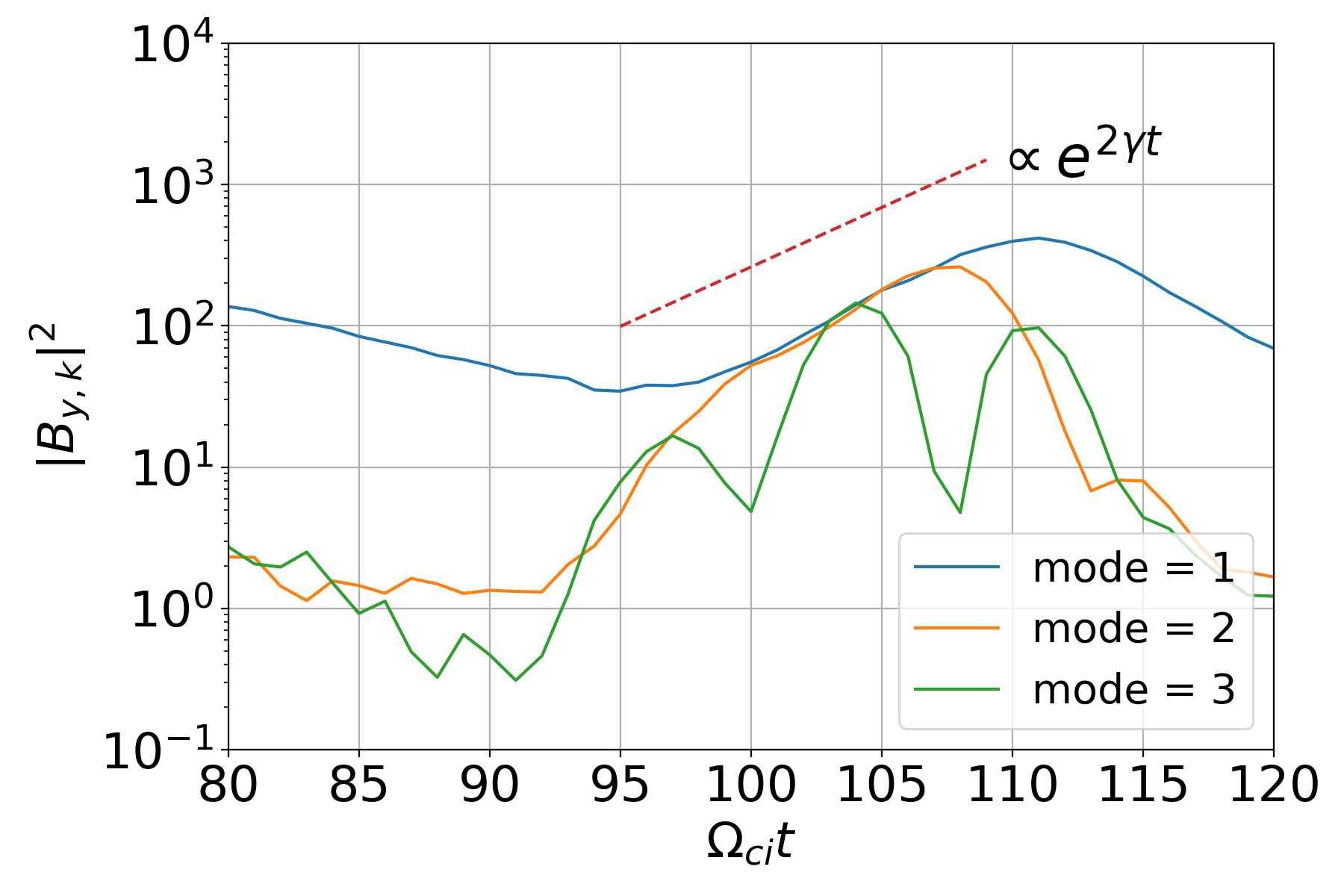}
    \caption{\label{chap3:compare_to_tearing}Comparison to linear growth rate \eqref{chap3:growth_rate_eq_max} and the Fourier component of $B_y$ in $x = [40 \lambda_i, 60 \lambda_i], y = 0.0$ for $m_i / m_e = 2$ (Run 2) case when several secondary plasmoids exist in the electron diffusion region. Mode 2 is well consistent with Eq.~\eqref{chap3:growth_rate_eq_max} after secondary plasmoids are beginning to appear around $90 \Omega_{ci}^{-1}$.}
\end{figure}

Figure \ref{chap3:de_with_plasmoid} presents the ion number density and the electron diffusion region\cite{zenitani2011prl, zenitani2011pop} during the formation of secondary plasmoids for $m_i / m_e = 2$ case. The electron diffusion region is identified as the area where $D_e := j_\mu E^\mu > 0$ in the electron rest frame\cite{zenitani2011prl}. Here $j_\mu := (\rho_q, \mathbf{j})$ and $E^\mu := F^{\mu \nu} u_\nu$ where $\rho_q$ is the charge density, $\mathbf{j}$ is the current, $F^{\mu\nu}$ is the electromagnetic tensor and $u_\nu$ is the four-velocity. $D_e$ can also be written as $D_e = \gamma_e (\mathbf{j} \cdot (\mathbf{E + v_e \times B}) - \rho_q (\mathbf{v_e \cdot E}))$, where $\gamma_e$ is the Lorentz factor of electron flow. The figure suggests that secondary plasmoids form within this region.

To investigate the mechanism of secondary plasmoid formation, we focus on the electron diffusion region and compare it with the growth rate of collisionless electron tearing instability. The linear growth rate is expressed as:
\begin{equation}
    \begin{aligned}
        \gamma & = \frac{k v_{Te}}{\sqrt{\pi}} \left( \frac{u_e}{v_{Te}} \right)^{\frac{3}{2}} \frac{1 + T_i / T_e}{1 + (m_e T_e / m_i T_i)^{1/4}}
(1 - k^2 \delta^2),
    \end{aligned}
    \label{chap3:growth_rate_eq}
\end{equation}
where $v_{Te}$ is the electron thermal speed, $u_e$ is the out-of-plane electron bulk speed at the current sheet, and $k$ is the wave number of the growth mode\cite{zelenyi1979}. The maximum growth rate is given by:
\begin{equation}
    \begin{aligned}
        k_{\text{max}} &= \frac{1}{\sqrt{3} \delta}, \\
        \gamma_{\text{max}} &= \frac{2 \sqrt{3}}{9 \delta} \frac{v_{Te}}{\sqrt{\pi}} \left( \frac{u_e}{v_{Te}} \right)^{\frac{3}{2}} \frac{1 + T_i / T_e}{1 + (m_e T_e / m_i T_i)^{1/4}}.
    \end{aligned}
    \label{chap3:growth_rate_eq_max}
\end{equation}

Figure \ref{chap3:compare_to_tearing} compares the growth rate of the Fourier component for the $m_i / m_e = 2$ case with the linear growth rate from Eq.~\eqref{chap3:growth_rate_eq_max}. In the simulation, the growth rate is computed using the $B_y$ component within $x = [40 \lambda_i, 60 \lambda_i]$. The linear growth rate is derived from the background electron temperature, $v_{Te} = \sqrt{2 T_{eb} / m_e}$, and the simulated out-of-plane electron bulk velocity, $u_e$. The observed modes at $t = 95 \sim 100 \Omega_{ci}^{-1}$ align well with the linear growth rate predicted by Eq.~\eqref{chap3:growth_rate_eq_max}, indicating that secondary plasmoid formation is driven by collisionless electron tearing instability inside the electron diffusion region.

\begin{figure}[htbp]
    \includegraphics[width=\linewidth]{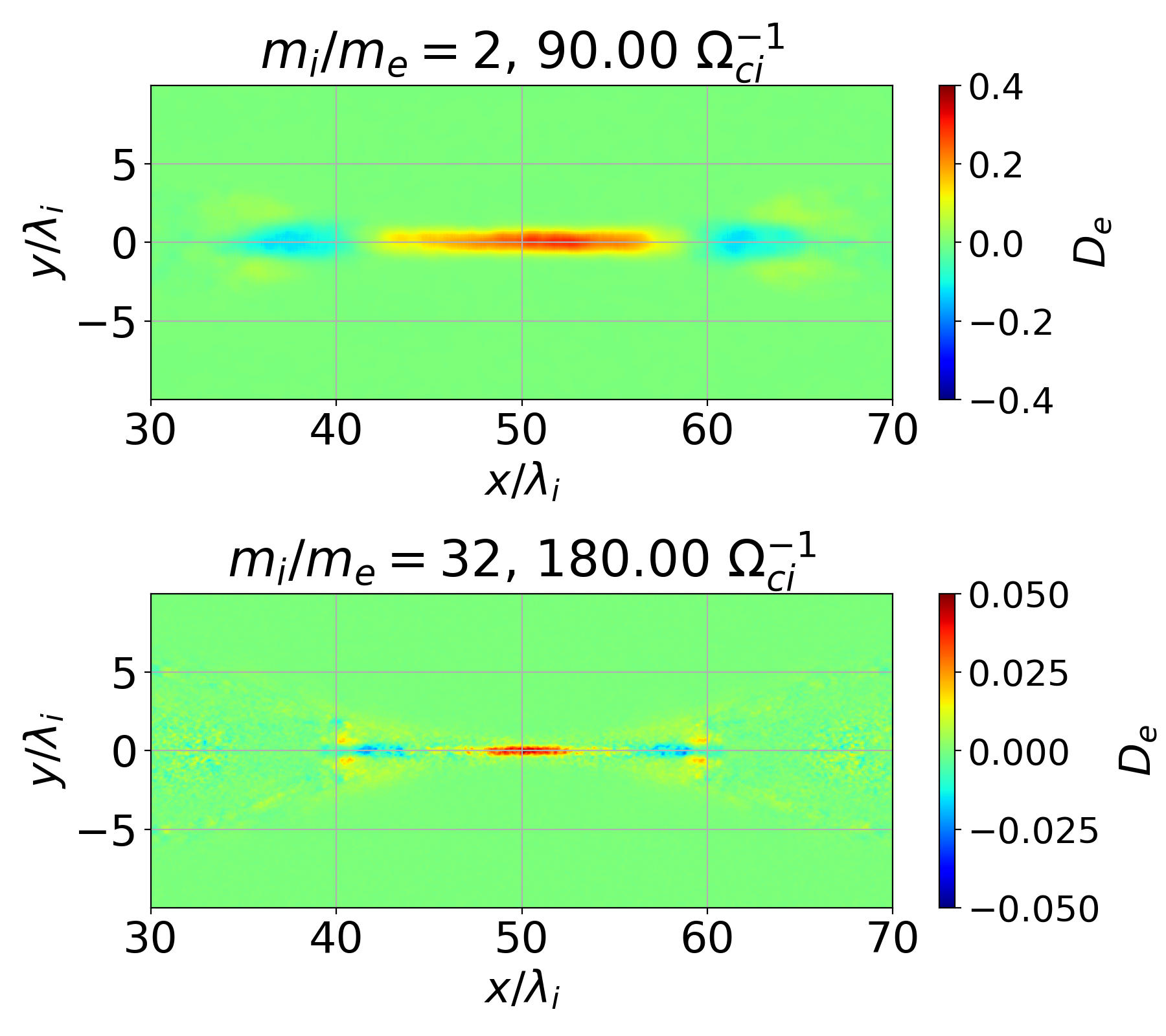}
    \caption{\label{chap3:de_different_mass_ratio}Spatial distribution of $D_e$ for $m_i / m_e = 2, 32$ cases. Red area ($D_e > 0$) corresponds to the electron diffusion region.}
\end{figure}

\begin{figure}[htbp]
    \includegraphics[width=\linewidth]{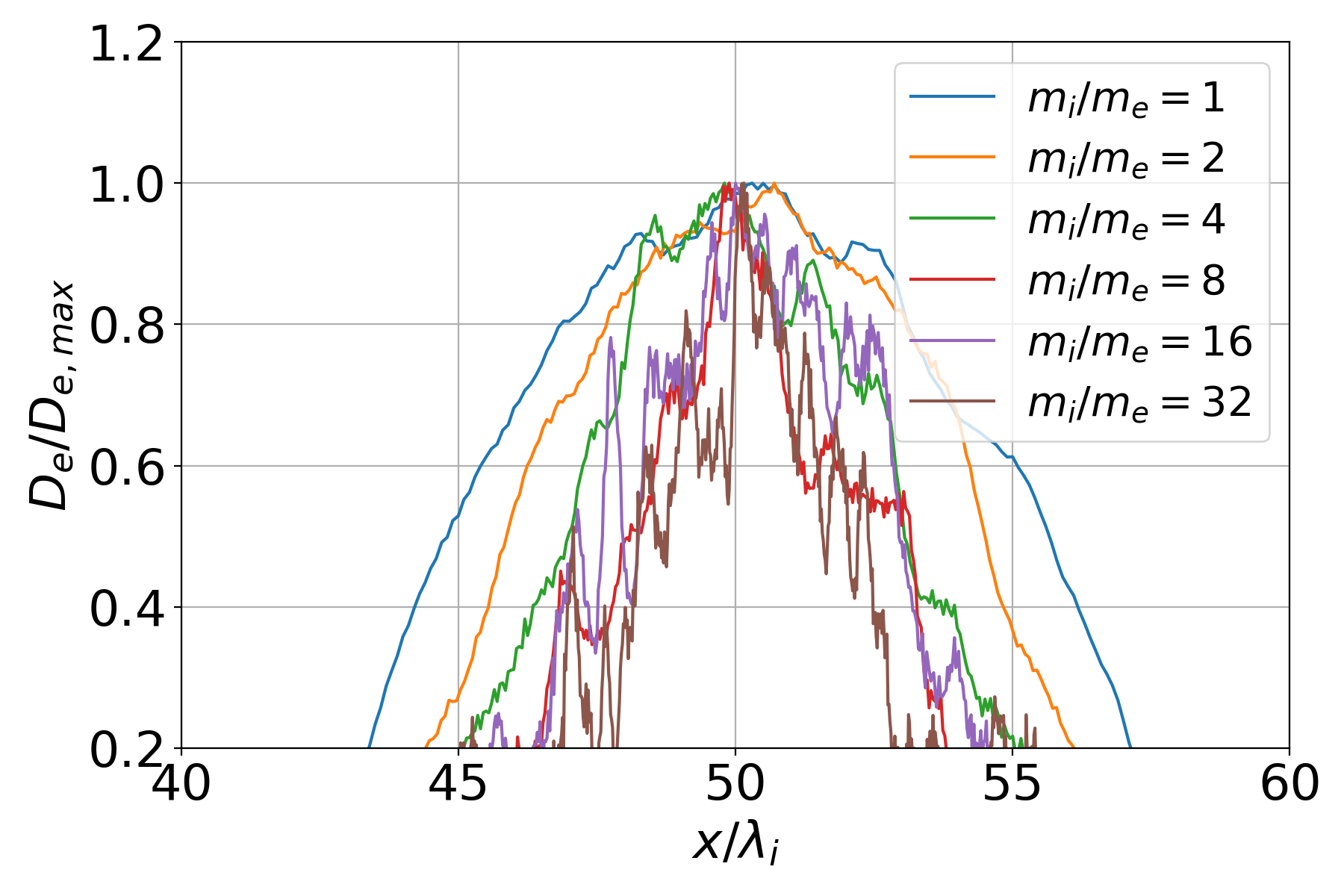}
    \caption{\label{chap3:de_different_mass_ratio_1dcut}Spatial distribution of $D_e / D_{e, max}$ cut in $y = 0$ for each $m_i / m_e$ case. This figure shows $D_e / D_{e, max} > 0.2$ where we define as the effective electron diffusion region. The length is normalized by ion inertial length $\lambda_i$ for each $m_i / m_e$.}
\end{figure}

To illustrate the dependence of the electron diffusion region on the mass ratio, Figure \ref{chap3:de_different_mass_ratio} presents $D_e$ for $m_i / m_e = 2, 32$. As $m_i / m_e$ increases, the size of the electron diffusion region ($D_e > 0$ area) normalized by $\lambda_i$ decreases. The value of $D_e$ also declines with increasing $m_i / m_e$. Figure \ref{chap3:de_different_mass_ratio_1dcut} displays the spatial distribution of $D_e / D_{e, {\rm max}}$ along $y = 0$. Here, we define the effective electron diffusion region as the area where $D_e / D_{e, {\rm max}} > 0.2$. \rewrite{Changing the threshold does not affect the $m_i / m_e$ dependence of the length of the electron diffusion region.} The length of this region, normalized by $\lambda_i$, shrinks as $m_i / m_e$ increases.


\subsection{\label{chap3:x_point_around_secondary_plasmoid}X point around secondary plasmoid}

\begin{figure}[htbp]
    \includegraphics[width=\linewidth]{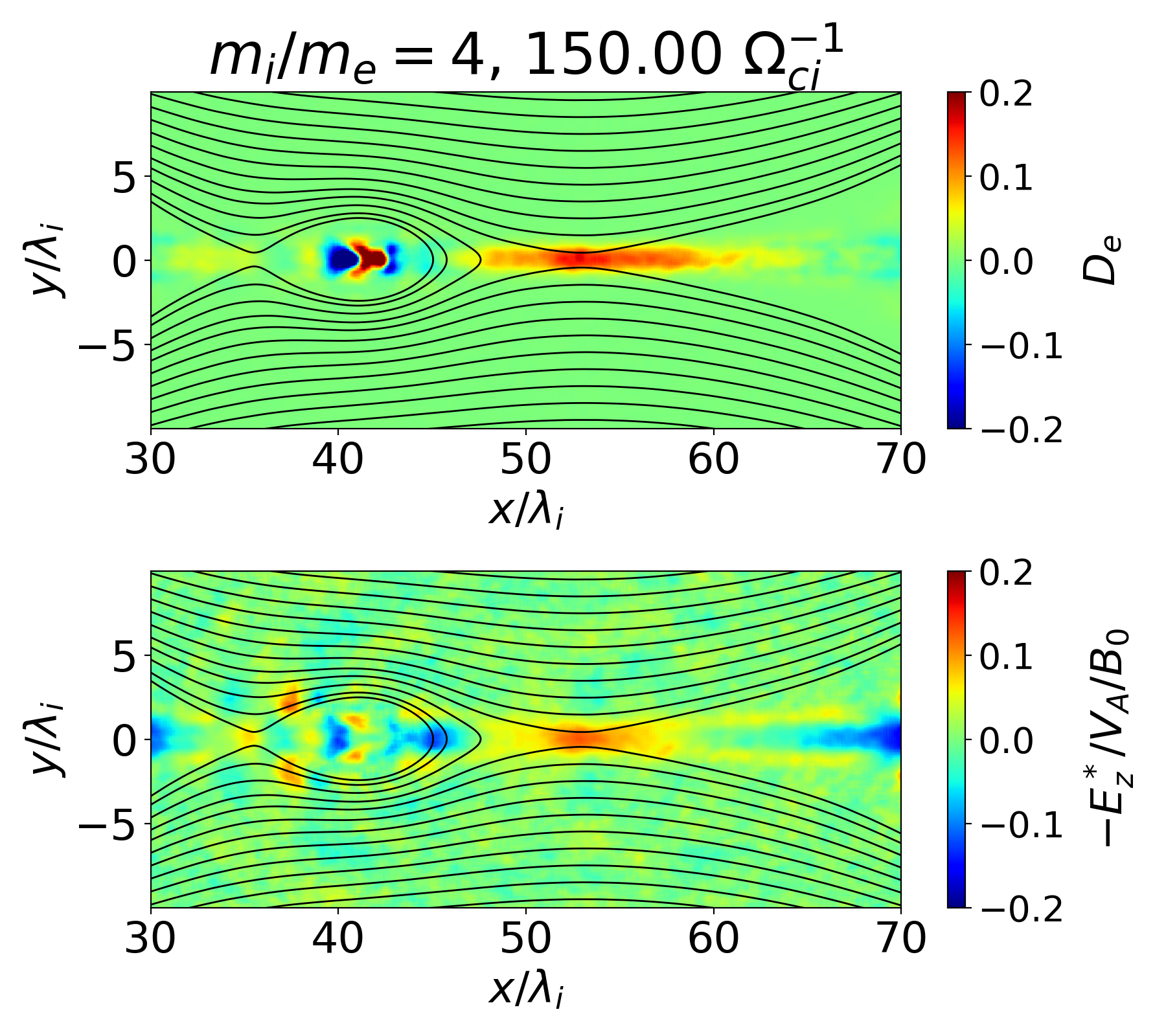}
    \caption{\label{chap3:de_ez_with_plasmoid}$D_e$ and negative sign of non-ideal electric field $-E_z^* = -(\bm{E} + \bm{v_e} \times \bm{B})_z$ when secondary plasmoid exists for $m_i / m_e = 4$ case. Black line shows the in-plane magnetic field lines.}
\end{figure}

To illustrate the nature of magnetic field dissipation near the X-point, Figure \ref{chap3:de_ez_with_plasmoid} presents $D_e$ and the negative of the non-ideal out-of-plane electric field, $-E_z^* = -(\bm{E} + \bm{v_e} \times \bm{B})_z$. These figures indicate that the region surrounding the X-point associated with the secondary plasmoid ($x \sim [30 \lambda _i, 40 \lambda _i]$, hereafter referred to as the "secondary plasmoid X-point") exhibits no magnetic dissipation ($D_e \sim 0$) in contrast to the electron diffusion region, where dissipation is present ($D_e > 0$ at $x \sim [50 \lambda _i, 60 \lambda _i]$).

\begin{figure}[htbp]
    \includegraphics[width=\linewidth]{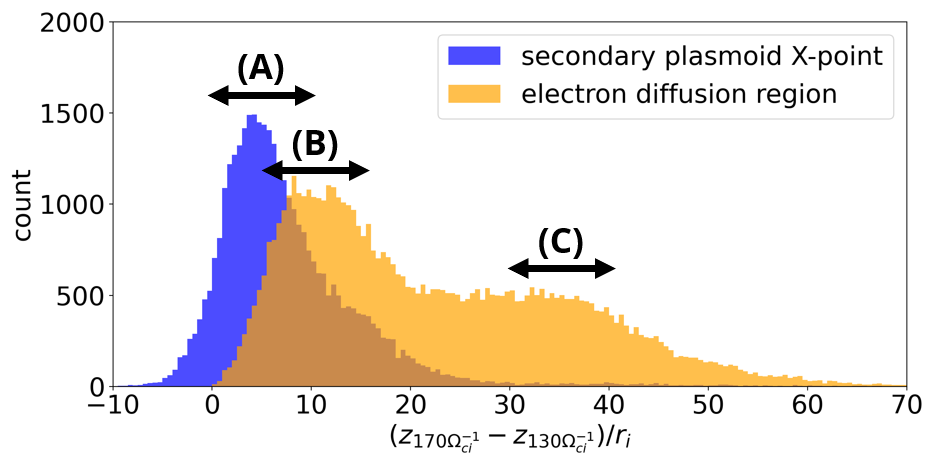}
    \caption{\label{chap3:moved_distance_distribution}Histograms of moved distance of electron in the z direction for $m_i / m_e = 4$ case. Blue histogram represents the electrons which pass secondary plasmoid X-point ($x \sim [30\lambda_i, 40\lambda_i], y \sim [-0.5\lambda_i, 0.5\lambda_i]$) at $150 \Omega_{ci}^{-1}$. Orange histogram represents the electrons which pass the electron diffusion region ($x \sim [50\lambda_i, 60\lambda_i], y \sim [-0.5\lambda_i, 0.5\lambda_i]$) at $150 \Omega_{ci}^{-1}$. Moved distance is calculated by $z(t = 170 \Omega_{ci}^{-1}) - z(t = 130 \Omega_{ci}^{-1})$. }
\end{figure}

\begin{figure}[htbp]
    \includegraphics[width=\linewidth]{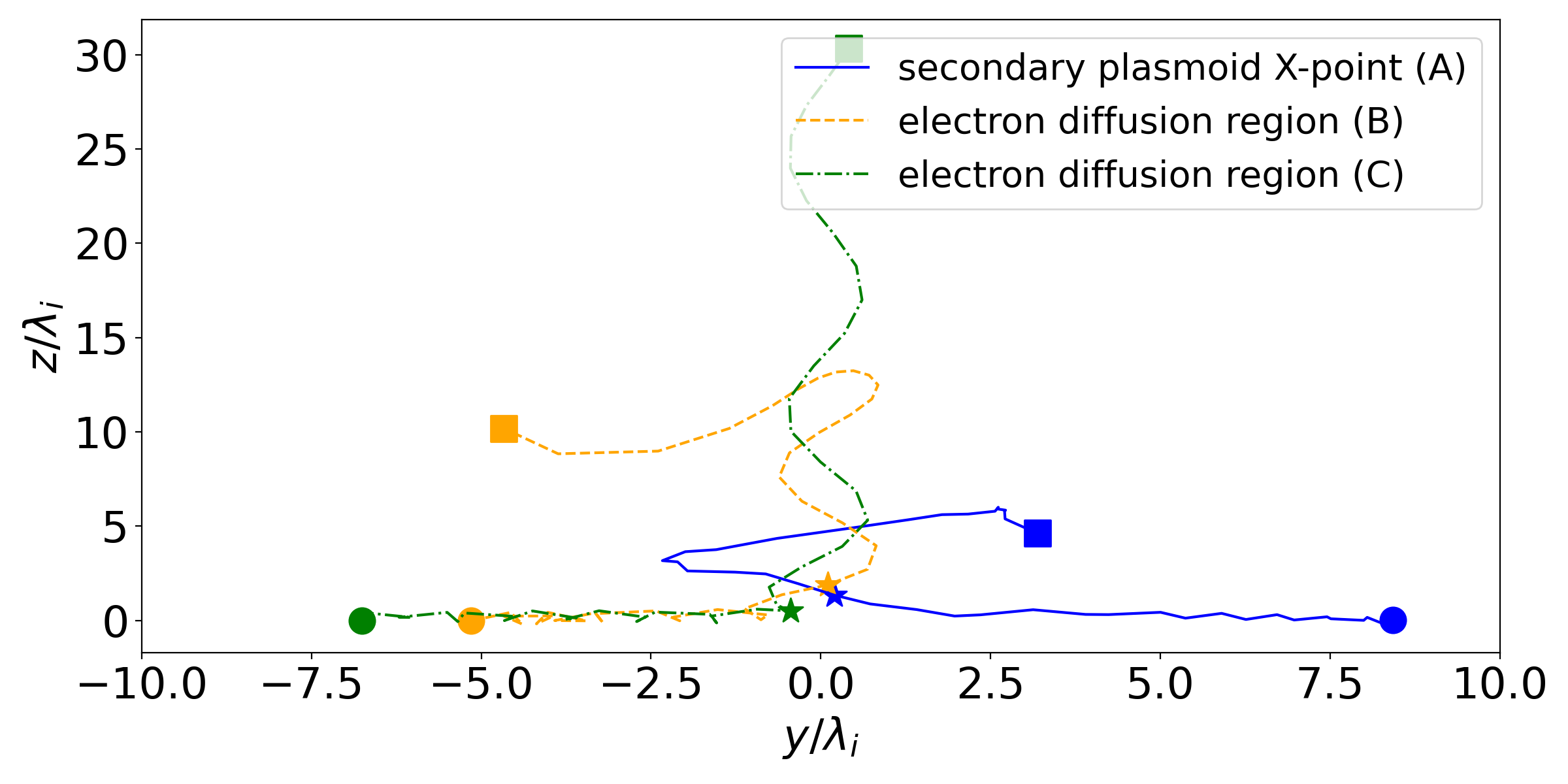}
    \caption{\label{chap3:particle_orbit}Orbits of random sampled particle around electron diffusion region and secondary plasmoid X-point. The regions of (A), (B) and (C) to pick up one sample are shown in Figure \ref{chap3:moved_distance_distribution}. \rewrite{The circle shows the start position at $t = 130\Omega_{ci}^{-1}$ and the square shows the end position at $t = 170 \Omega_{ci}^{-1}$}}
\end{figure}

\begin{figure}[htbp]
    \includegraphics[width=0.95\linewidth]{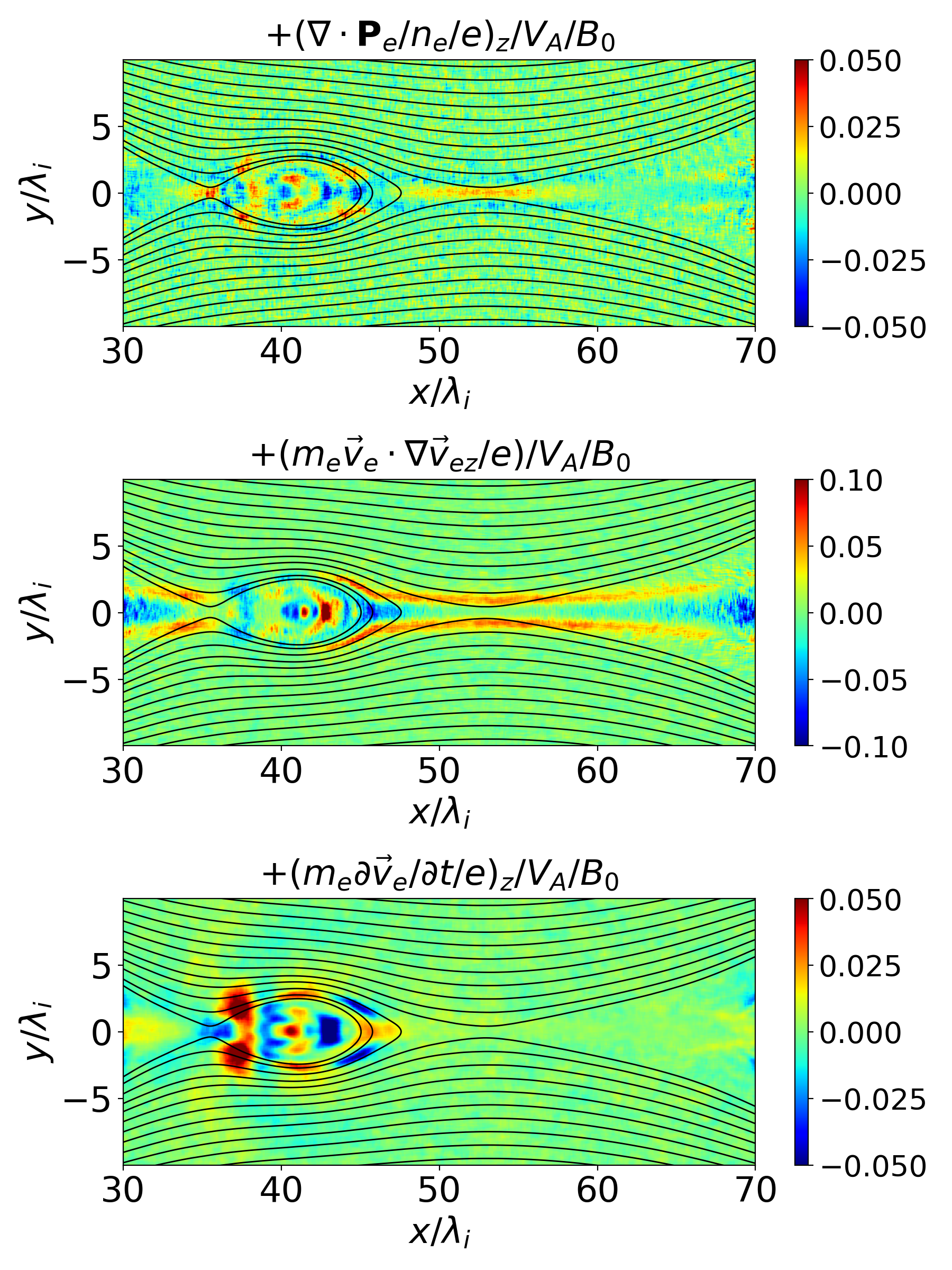}
    \caption{\label{chap3:ez_components}Each component of out-of-plane non-ideal electric field $E_z$ for $m_i / m_e = 4$ case (see Eq.~\eqref{chap3:ez_component_eq}). Black line shows the in-plane magnetic field lines.}
\end{figure}

To clarify the origin of minimal dissipation \rewrite{near the secondary plasmoid X-point region}, we compare particles in the electron diffusion region and the secondary plasmoid X-point region. Figure \ref{chap3:moved_distance_distribution} presents histograms of the displacement in the z direction for electrons passing through each region at $150 \Omega_{ci}^{-1}$. Both histograms exhibit a single peak, but the electron diffusion region shows an excess in the range of approximately $20 \sim 40$ on the horizontal axis. This indicates distinct electron behaviors in the two regions. 

To explore this further, we conduct a particle orbit analysis. Figure \ref{chap3:particle_orbit} presents randomly sampled orbits corresponding to the peak of the secondary plasmoid X-point region (A, straight line), the peak of the electron diffusion region (B, dotted line), and \rewrite{the flatten area where particles moved relatively large in the z direction through the electron diffusion region} (C, dashed line). The orbits at (A) and (B) suggest the cyclotron motion, as they almost do not move in the z direction for about half or one gyration period. In contrast, the orbit in area (C) exhibits the meandering motion, oscillating around $y = 0$ and moving in z direction. Meandering particles can resonate with $E_z^*$\rewrite{, receiving energy from magnetic field effectively.} In conclusion, almost no meandering particles are observed in the secondary plasmoid X-point region.

To explore the physical origin of the nearly-zero $E_z^*$ at the secondary plasmoid X-point, we plot the components of $E_z^*$. From the electron equation of motion, $E_z^*$ can be expressed as:
\begin{equation} 
    \begin{aligned} 
        E_z^* &= \left( \frac{\nabla \cdot \bm{P_e}}{n_e e} + \frac{m_e}{e} \frac{d\bm{v_e}}{dt} \right)_z \\ &= \frac{\frac{\partial P{xze}}{\partial x} + \frac{\partial P_{yze}}{\partial y}}{n_e e} + \frac{m_e}{e} \left( \frac{\partial v_{ez}}{\partial t} + \bm{v_e} \cdot \nabla v_{ez} \right). \label{chap3:ez_component_eq}
    \end{aligned}
\end{equation}

Figure \ref{chap3:ez_components} presents the components of \eqref{chap3:ez_component_eq}, referred to as the electron pressure term and the electron inertial term. These figures indicate that the electron pressure term is the main contributor to $E_z^*$ inside the electron diffusion region. A previous study also examined these components to determine the origin of $E_z^*$ and concluded that the electron pressure term is dominant\cite{hesse1999}. The electron pressure gradient arises from meandering particles, which is consistent with Figure \ref{chap3:moved_distance_distribution}. \rewrite{On the other hand, the electron pressure term and the electron inertial term are canceled at secondary plasmoid X-point. It indicates that meandering particles are swept out from secondary plasmoid X-point due to the motion of secondary plasmoid in the x direction and $E_z^*$ becomes nearly zero. It should be the reason why the histogram in Figure \ref{chap3:moved_distance_distribution} shows that almost no particles are meandering around secondary plasmoid X-point.}

\section{\label{chap4}Comparison to plasmoid instability}

As discussed in the previous section, secondary plasmoids appear in some PIC simulations with a small ion-electron mass ratio $m_i / m_e$, which look like plasmoid instability in resistive MHD simulations. However, this does not necessarily imply identical underlying physics; secondary plasmoids in PIC simulations play the same roles as the plasmoid instability. Furthermore, the difference between the absence of secondary plasmoids in high $m_i / m_e$ PIC simulation and plasmoid instability in resistive MHD simulation is unclear. To address this concern, we present a comparative analysis of secondary plasmoids in PIC and MHD simulations in this section.

\subsection{\label{chap4:diffusion_region}Diffusion region}

We begin by comparing the properties of the current sheet and the diffusion region. To this end, we examine two PIC simulation results (Run 3B and 5B) and one MHD simulation result (Run 7).

\begin{figure}[htbp]
    \centering
    \includegraphics[width=\linewidth]{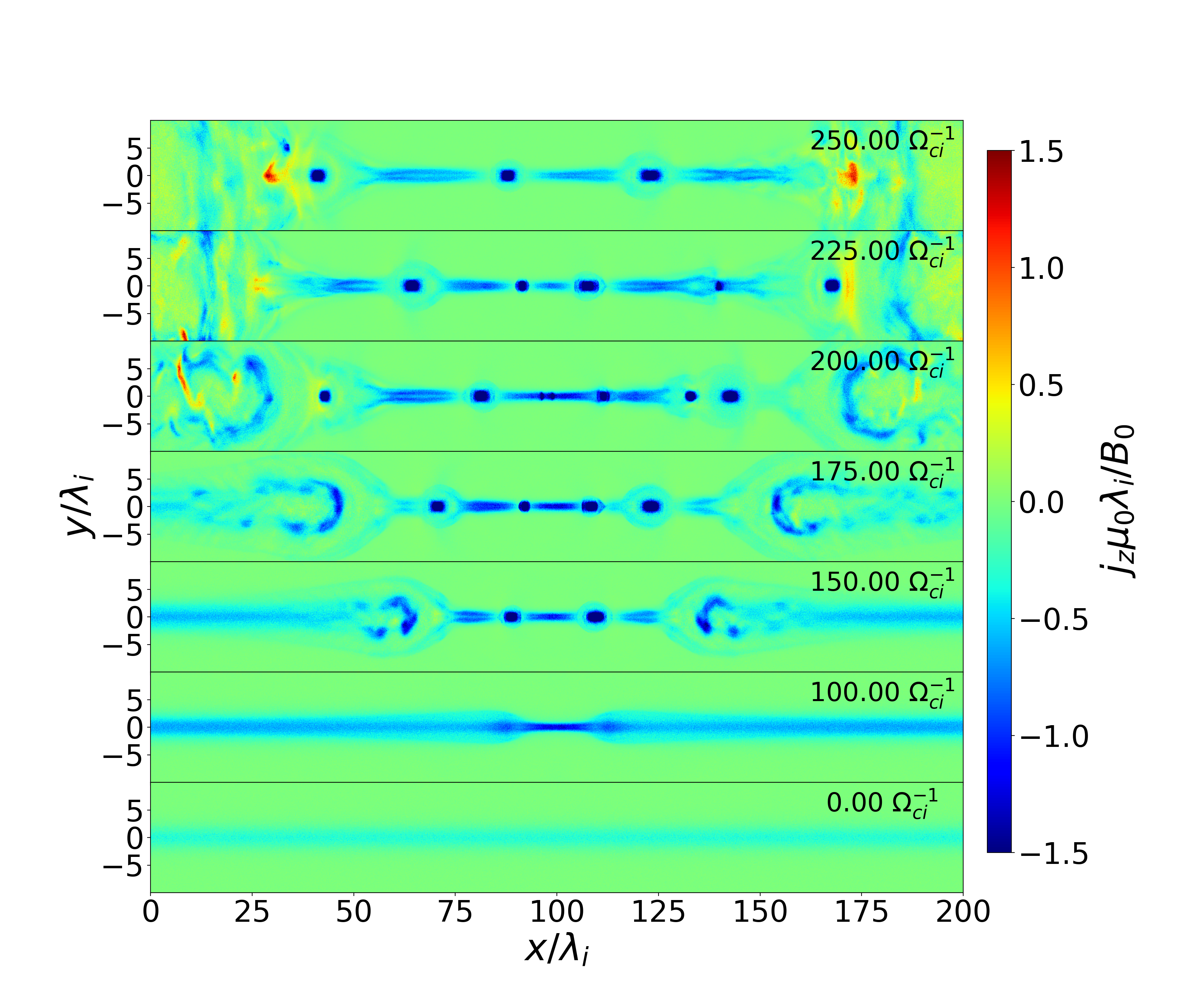}
    \caption{\label{chap4:time_evolution_jz_mr4}Time evolution of $j_z$ for Run 3B case.}
\end{figure}

\begin{figure}[htbp]
    \centering
    \includegraphics[width=\linewidth]{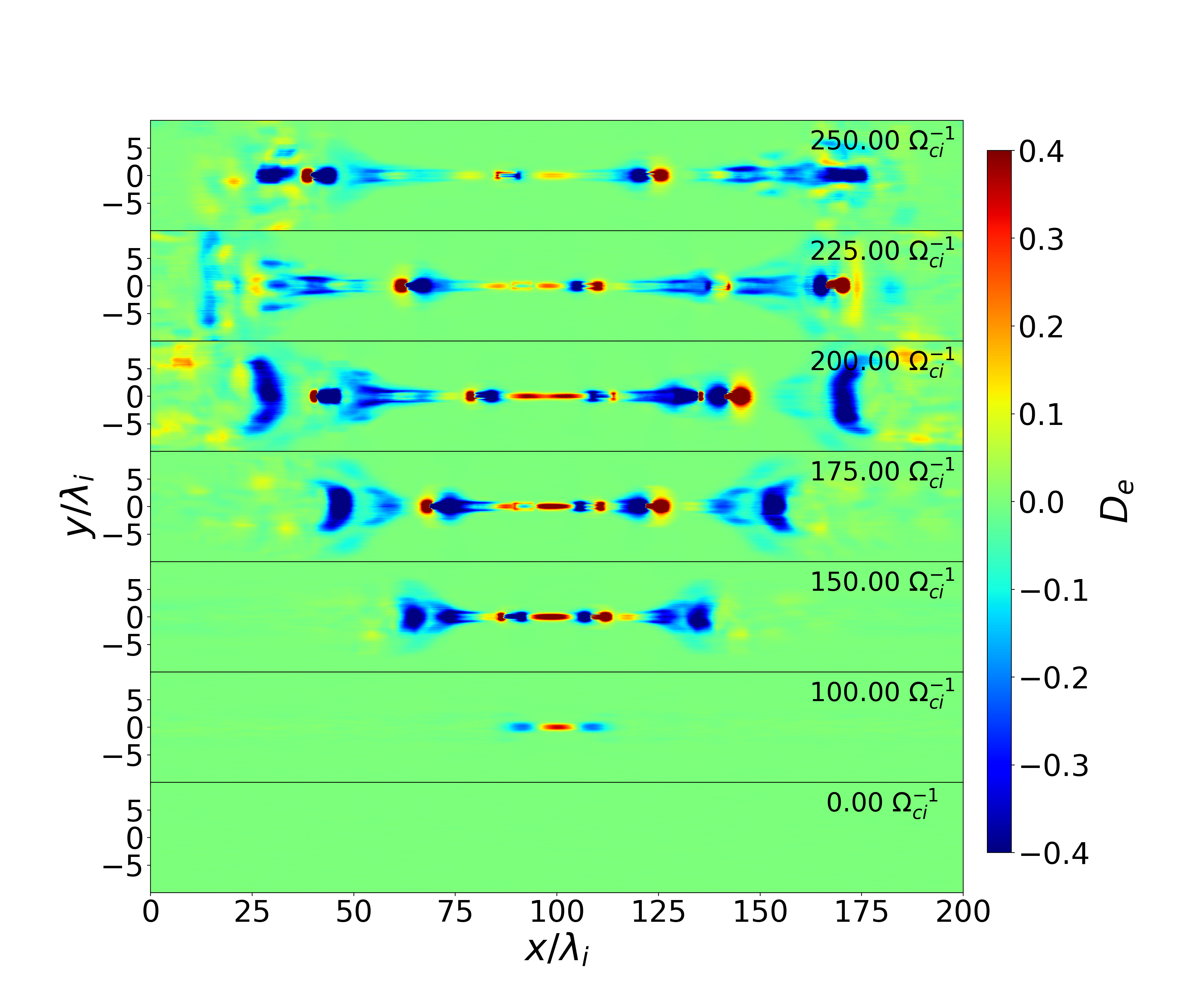}
    \caption{\label{chap4:time_evolution_de_mr4}Time evolution of the $D_e := j_\mu E^\mu$ for Run 3B case.}
\end{figure}

Figure \ref{chap4:time_evolution_jz_mr4} and \ref{chap4:time_evolution_de_mr4} present the results of the out-of-plane current $j_z$ and $D_e := j_\mu E^\mu$ for Run 3B case. Due to the small mass ratio $m_i / m_e$, secondary plasmoids are generated from the current sheet during magnetic reconnection. Figure \ref{chap4:time_evolution_jz_mr4} shows that an elongated current sheet forms at $t \sim 100 \Omega_{ci}^{-1}$, followed by the generation of multiple secondary plasmoids after the onset of magnetic reconnection. After $t = 150 \Omega_{ci}^{-1}$, plasmoid formation becomes non-steady. Figure \ref{chap4:time_evolution_de_mr4} illustrates that secondary plasmoids emerge inside the electron diffusion region and are subsequently advected. Pairs of $D_e > 0$ and $D_e < 0$ are observed to flow together. Here, $D_e > 0$ corresponds to the electron diffusion region, while $D_e < 0$ represents the region where plasma energy is converted into electromagnetic energy, indicating the presence of a dynamo effect. This suggests that dissipation and the dynamo effect occur simultaneously, resulting in negligible net dissipation. In Section \ref{chap3}, particle orbit analysis confirms that magnetic dissipation does not occur near the X-point of secondary plasmoids. This figure further supports that no significant magnetic dissipation occurs overall.

\begin{figure}[htbp]
    \centering
    \includegraphics[width=\linewidth]{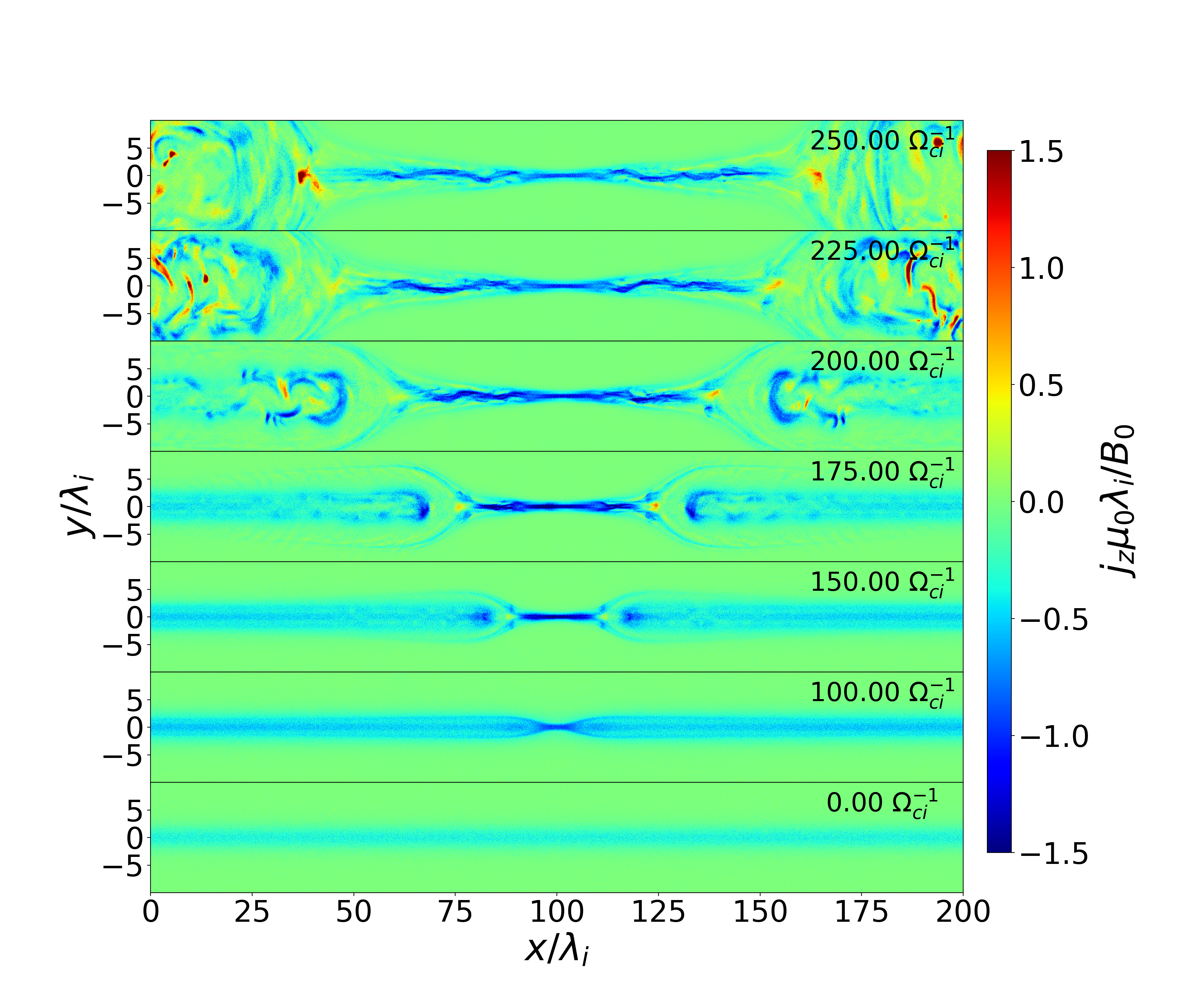}
    \caption{\label{chap4:time_evolution_jz_mr16}Time evolution of $j_z$ for Run 5B case.}
\end{figure}

\begin{figure}[htbp]
    \centering
    \includegraphics[width=\linewidth]{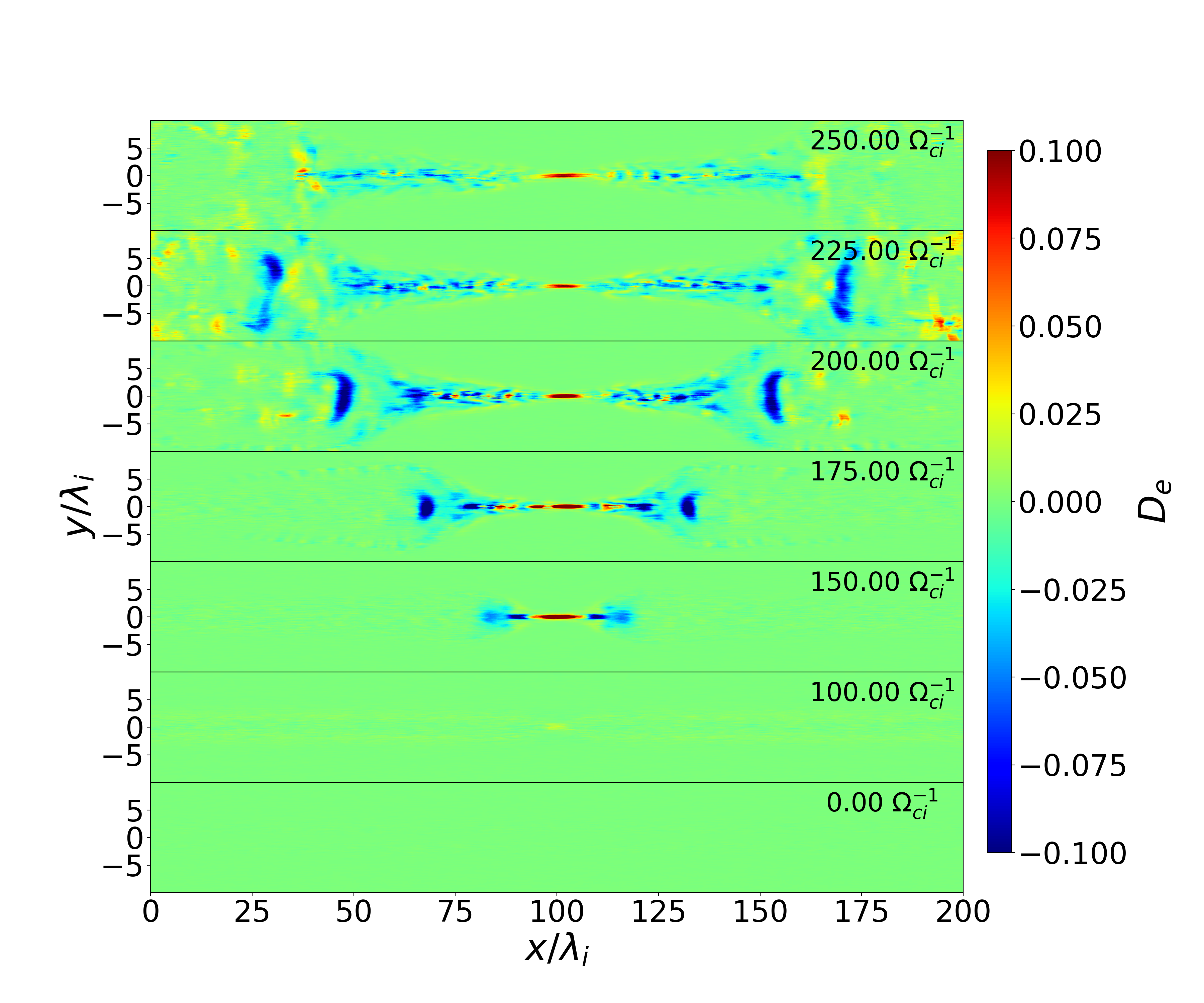}
    \caption{\label{chap4:time_evolution_de_mr16}Time evolution of the $D_e := j_\mu E^\mu$ for Run 5B case.}
\end{figure}

Figure \ref{chap4:time_evolution_jz_mr16} and \ref{chap4:time_evolution_de_mr16} present the results of the out-of-plane current $j_z$ and $D_e := j_\mu E^\mu$ in Run 5B. Figure \ref{chap4:time_evolution_jz_mr16} illustrates $j_z$. Similar to Run 3B case, an elongated current sheet forms in the early stage, but its subsequent evolution differs significantly. No secondary plasmoids appear. Around $t \sim 200 \Omega_{ci}^{-1}$, fluctuations begin in the current sheet. Moreover, the magnitude of $j_z$ at $t = 250 \Omega_{ci}^{-1}$ is smaller than at $t = 150 \Omega_{ci}^{-1}$, indicating an increase in sheet thickness and stabilization against collisionless tearing instability. Figure \ref{chap4:time_evolution_de_mr16} shows that the electron diffusion region remains at the center, with its size unchanged over time. As discussed in Section \ref{chap3}, when $m_i / m_e = 16$, the size is approximately $20 \lambda_e = 5 \lambda_i$, where $\lambda_e$ and $\lambda_i$ are the electron and ion inertial lengths, respectively. Outside this region, areas with $D_e < 0$ appear, indicating the presence of a dynamo effect.

\begin{figure}[htbp]
    \centering
    \includegraphics[width=\linewidth]{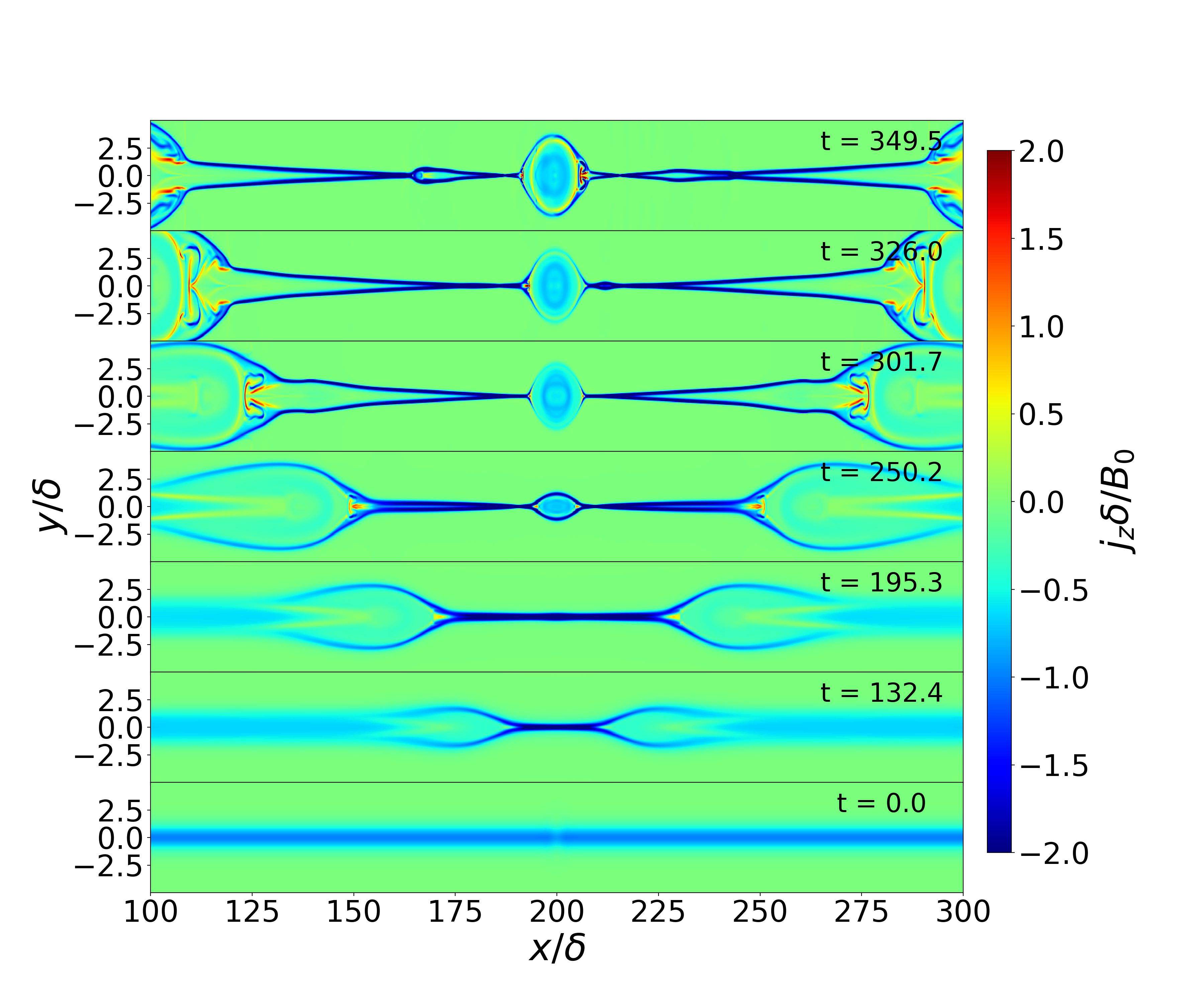}
    \caption{\label{chap4:time_evolution_jz_mhd}Time evolution of $j_z$ for Run 7 case.}
\end{figure}

\begin{figure}[htbp]
    \centering
    \includegraphics[width=\linewidth]{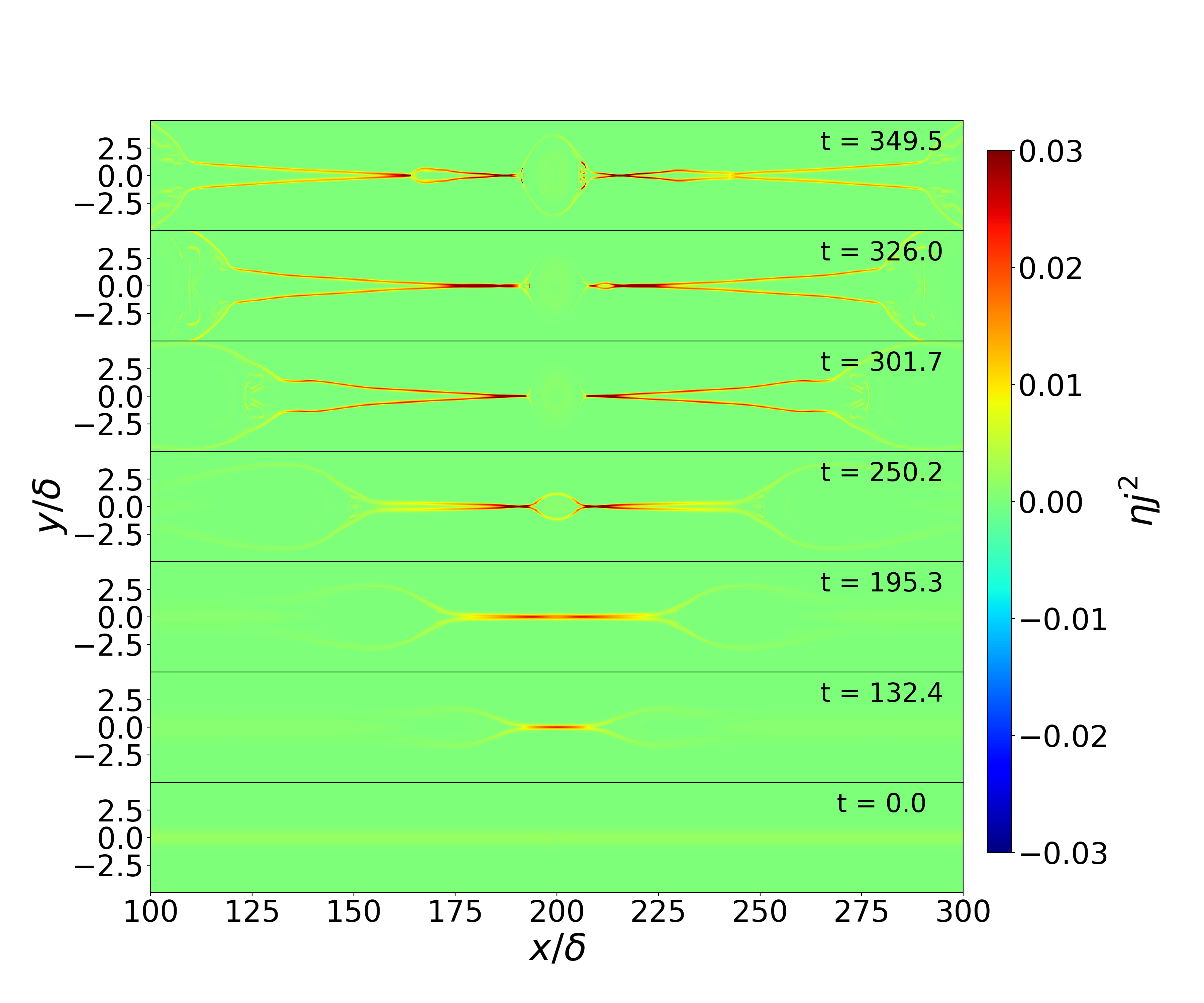}
    \caption{\label{chap4:time_evolution_etaj2_mhd}Time evolution of the $\eta j^2$ for Run 7 case.}
\end{figure}

Figure \ref{chap4:time_evolution_jz_mhd} and \ref{chap4:time_evolution_etaj2_mhd} present the time evolution of the out-of-plane current $j_z$ and Joule dissipation $\eta j^2$ for the resistive MHD case (Run 7). Until $t V_A / \delta \sim 200$, the diffusion region elongates horizontally, indicating that the reconnecting current sheet forms a structure similar to the Sweet-Parker current sheet. At $t V_A / \delta \sim 250$, plasmoid instability generates two localized diffusion regions. Additionally, Joule dissipation increases at the two X-points formed by plasmoid instability. By $t V_A / \delta \sim 320$, the dissipation region elongates again, and at $t V_A / \delta \sim 350$, multiple small diffusion regions emerge.

\subsection{\label{chap4:energy_conversion}Energy conversion}

Energy conversion efficiency or reconnection rate, alongside the current sheet structure, is an important aspect of magnetic reconnection. To this end, we examine PIC simulation result (Run 3A) and MHD simulation result (Run 7) to assess notable differences.

\begin{figure}[htbp]
    \centering
    \includegraphics[width=\linewidth]{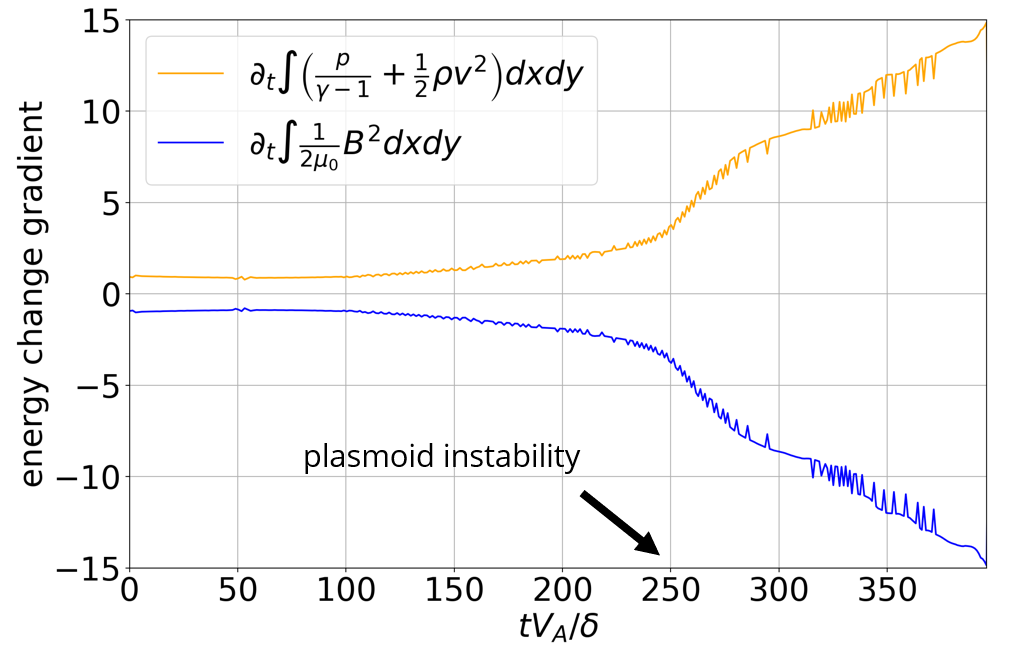}
    \caption{\label{chap4:time_evolution_energy_change_gradient}Time evolution of the energy change gradient of the kinetic and magnetic energy in the resistive MHD case (Run 7). Each energy is calculated as the sum of the entire simulation box.}
\end{figure}

The energy conversion efficiency is evaluated using the time derivatives of the magnetic and plasma (thermal + kinetic) energies, integrated over the entire simulation domain, as shown in Figure \ref{chap4:time_evolution_energy_change_gradient}. The rate of change increases at $t V_A / \delta \sim 250$, corresponding to the onset of the plasmoid instability, indicating that plasmoid instability accelerates energy conversion in the system.

\begin{figure}[htbp]
    \centering
    \includegraphics[width=\linewidth]{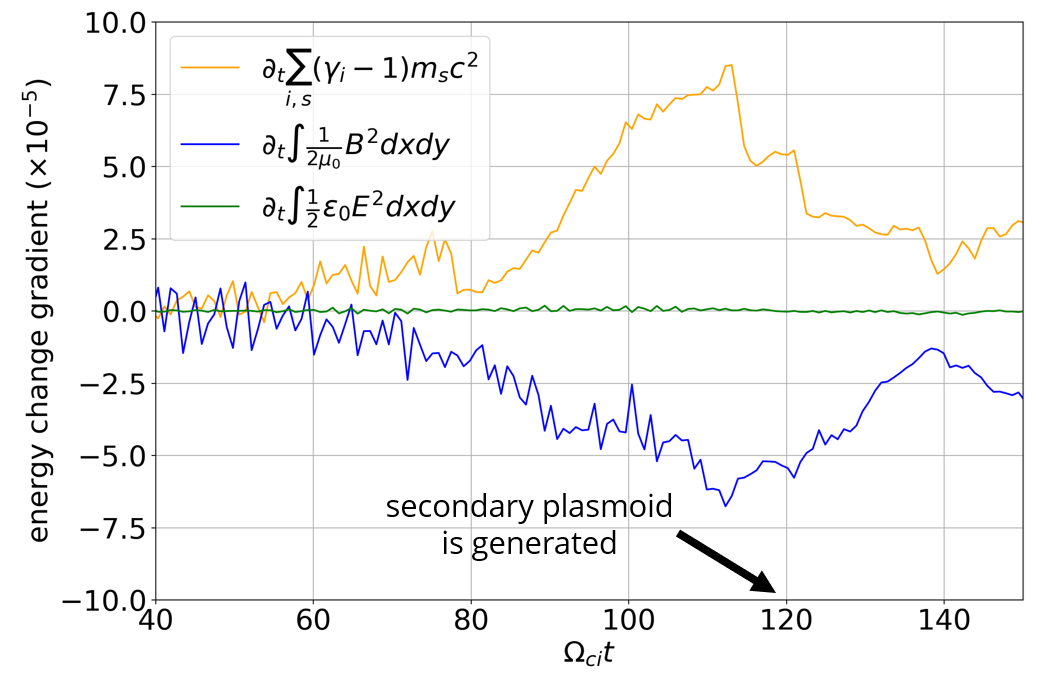}
    \caption{\label{chap4:pic_mr4_energy}Time evolution of the energy change gradient of the total kinetic energy, magnetic field energy and electric field energy \rewrite{in the PIC case (Run 3A).} Each energy is calculated as the sum of the entire simulation box.}
\end{figure}

Meanwhile, PIC simulations show a different behavior. \rewrite{Figure \ref{chap4:pic_mr4_energy} displays the time evolution of the total kinetic energy ($\sum_{i, s} (\gamma_i - 1) m_s c^2$), magnetic energy ($\int B^2 / 2\mu_0 dx dy$) and electric energy ($\int \epsilon_0 E^2 / 2 dx dy$). Here, $i$ and $s$ denote particle index and species, respectively, and the integrals are over the entire simulation domain.} These results show that the efficiency of energy conversion increases after the onset of reconnection. However, it drops around $t \sim 120 \Omega_{ci}^{-1}$, coinciding with the formation of secondary plasmoids. It suggests that secondary plasmoid reduces the energy conversion efficiency. It is the opposite of plasmoid instabiltiy which increases the energy conversion efficiency.

\section{\label{chap5:discussion_and_conclusion}Summary and Discussion}

\begin{figure*}[htbp]
    \centering
    \includegraphics[width=\textwidth]{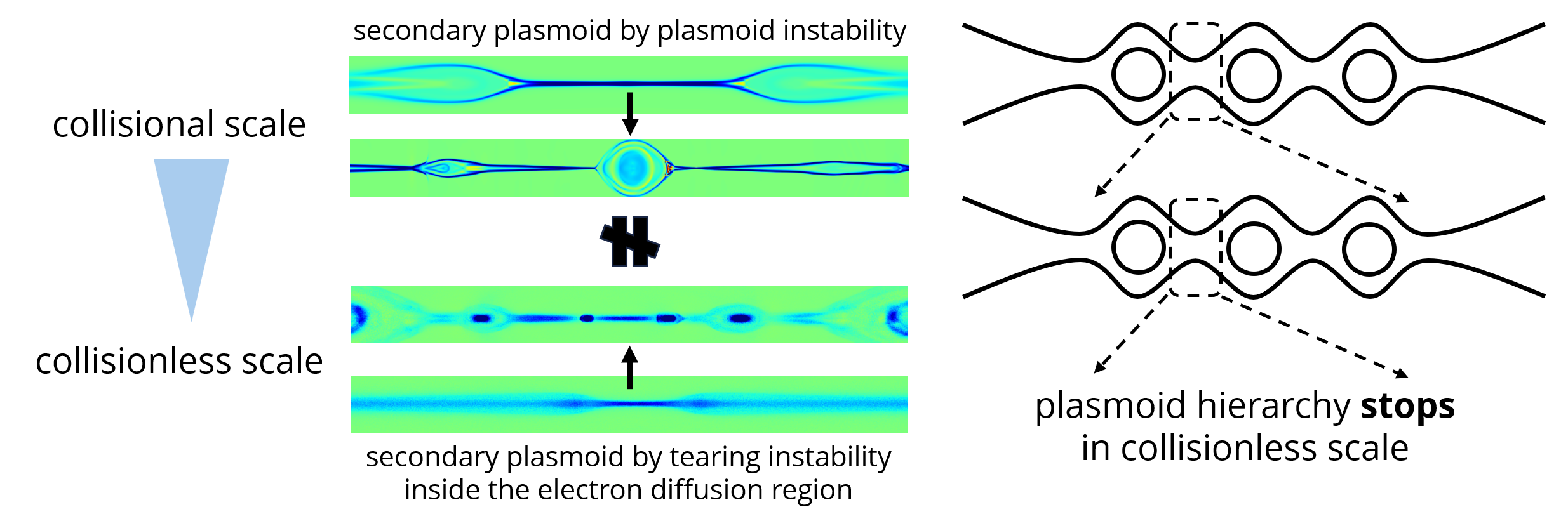}
    \caption{\label{chap5:conclusion_fig}Secondary plasmoids do not support plasmoid-mediated reconnection in \rewritesecond{2D antiparallel reconnection with higher mass ratio}.}
\end{figure*}

\rewrite{From the simulation results presented above, the following three findings have been clarified which are valid for 2D antiparallel collisionless magnetic reconnection within the parameters of our simulations:}
\begin{enumerate}
\item\label{chap5:result1} Secondary plasmoids form due to collisionless electron tearing instability inside the electron diffusion region. The electron diffusion region is on the electron scale, while plasmoid is on the ion scale. The key parameter controlling both scales is the mass ratio $m_i / m_e$ (Figure \ref{chap3:whole_results}, \ref{chap3:de_with_plasmoid}, \ref{chap3:compare_to_tearing}, \ref{chap3:de_different_mass_ratio}).
\item For low $m_i / m_e$ cases ($\lesssim 10$), secondary plasmoids appear; however, this is not due to plasmoid instability, 
as no enhancement of reconnection rate
(Figure \ref{chap4:time_evolution_jz_mr4}, \ref{chap4:time_evolution_de_mr4}, \ref{chap4:time_evolution_jz_mhd}, \ref{chap4:time_evolution_etaj2_mhd}, \ref{chap4:time_evolution_energy_change_gradient}, \ref{chap4:pic_mr4_energy}).
\item For high $m_i / m_e$ cases ($\gtrsim 10$), secondary plasmoids do not appear because the electron diffusion region does not elongate but remains constant in size (Figure \ref{chap4:time_evolution_jz_mr16}, \ref{chap4:time_evolution_de_mr16}, \ref{chap4:time_evolution_jz_mhd}, \ref{chap4:time_evolution_etaj2_mhd}).
\end{enumerate}
In particular, this study reveals that the role and formation mechanism of secondary plasmoids differ fundamentally between PIC and resistive MHD simulations. The key distinction lies in resistivity: while resistivity $\eta$ is externally imposed in MHD, it arises self-consistently in collisionless systems. \rewritesecond{Figure \ref{chap5:conclusion_fig} illustrates our conclusion that plasmoid-mediated reconnection do not occur in 2D antiparallel reconnection with higher mass ratio because the secondary plasmoids do not accelerate reconnection within physical parameters of our simulations.}

\rewrite{Some previous studies show that secondary plasmoids are formed in 2D reconnection with higher $m_i / m_e$ (2D case\cite{oka2010, lu2020}, 2D with guide field case\cite{drake2006grl, lu2020}, 3D case\cite{fujimoto2012}). Secondary plasmoid formations in antiparallel magnetic reconnection may be due to the coarse resolution. Comparison of PIC simulation results with different grid sizes are shown in Appendix A. To find out what happens with 2D reconnection with guide field or 3D reconnection is our future work. Moreover previous study\cite{almonte2014} reported that electrostatic instability occur inside the electron diffusion region when large $\omega_{pe} / \Omega_{ce}$ is used. To clarify the $\omega_{pe} / \Omega_{ce}$ dependence for secondary plasmoid formation and the size of the electron diffusion region is also our future work.}

In Section \ref{chap3}, we discuss the $m_i / m_e$ dependence of secondary plasmoid formation inside the electron diffusion region. The length of this region is estimated as
\begin{eqnarray} 
    L_{De} = 2 r^{-1} \lambda_e \sim 20 \lambda_e \ (r \sim 0.1) = 20 \sqrt{\left( \frac{m_e}{m_i} \right)} \lambda_i, \label{chap3:De_size}
\end{eqnarray}
where $\lambda_e$ is the electron inertial length and $r$ is the reconnection rate, typically $\mathcal{O}(0.1)$.
According to Eq.~\eqref{chap3:growth_rate_eq}, the current sheet becomes unstable for the tearing modes with wavelengths larger than the sheet thickness. Assuming the reconnecting current sheet thickness is $\mathcal{O}(1.0 \lambda_i)$, the instability condition becomes
\begin{eqnarray} 
    L_{De} \sim 20 \sqrt{\left( \frac{m_e}{m_i} \right)} \lambda_i > 2 \pi \lambda_i, \hspace{1em} \therefore \frac{m_i}{m_e} < \frac{100}{\pi^2} \sim 10. \label{chap3:unstable_condition} 
\end{eqnarray}
This threshold agrees with our results (Figure \ref{chap3:whole_results}) and implies that secondary plasmoids should not form when the real mass ratio ($\sim 1836$) is used.
\rewrite{Our results indicate that the reason why secondary plasmoid formation has $m_i / m_e$ dependence is that the ratio between the plasmoid size (ion scale) and the electron diffusion region size (electron scale) is controlled by $m_i / m_e$.}

In Section \ref{chap4}, we compare the current sheet structure and diffusion region. For large $m_i / m_e$ cases, the current sheet structure is almost the same between PIC and resistive MHD simulation results; however, secondary plasmoids rarely appear due to the fixed size of the electron diffusion region. The Hall effect becomes significant at higher $m_i / m_e$, generating an out-of-plane magnetic field (Hall magnetic field) that alters the reconnecting current sheet structure from the anti-parallel configuration. In contrast, an anti-parallel current sheet is formed in resistive MHD simulations. The Hall effect may partly explain why the plasmoid instability scarcely develops in collisionless systems. A detailed comparison with the Hall plasmoid instability reported in previous studies\cite{huang2011} is beyond the scope of this paper.

Several observational studies have reported secondary plasmoids in collisionless systems. For instance, secondary plasmoids have been observed in some solar flares\cite{takasao2012}. Solar flares are believed to be driven by hybrid magnetic reconnection involving both collisional and collisionless X-points\cite{ji2011}. In such events, a thin current sheet formed via plasmoid instability at the MHD scale can thin down to the kinetic scale (below $10^5$ m, or approximately $10^5 \lambda_i$, which corresponds to the electron-electron mean free path in the solar corona). Our simulation domain is $\mathcal{O}(10^2) \lambda_i$, indicating a large scale gap. It remains uncertain whether plasmoid instability occurs in systems of size $\mathcal{O}(10^5) \lambda_i$. On the other hand, since our simulation size is comparable to the Earth's magnetotail, our results imply that plasmoid instability should not appear there. This contradicts observations reporting persistent magnetic field variations caused by plasmoids from collisionless magnetic reconnection\cite{chen2008}.

\rewritesecond{In experimental study\cite{olson2016}, electron scale secondary plasmoids were found inside the diffusion region in collisionless systems. Note that our results indicate that plasmoid-mediated reconnection in 2D without guide field case may not occur and the situation is far from experimental one. However, we suggest that examining the reconnecting electric field in detail should be needed to discuss the difference between secondary plasmoid formation and plasmoid instability for both experimental and observational studies.}

Finally, future prospects are discussed. The system investigated in this study is of the order of $\mathcal{O}(10^2)\lambda_i$, while there is a scale gap of $10^3$ compared to the collisional scale in solar flares. In 2D collisionless magnetic reconnection simulations, extending the current sheet length by $10^3$ requires $10^6$ times more memory and (at least) $10^3$ times more computation time. With recent GPU advancements, this may become feasible. However, handling 3D effects or full-scale systems ($\mathcal{O}(10^{7 \sim 8})\lambda_i$) presents major challenges. We consider multi-hierarchy simulations a promising approach to this issue\cite{sugiyama2007, usami2013, daldorff2014, makwana2017}, and an important direction for future work.

\section{\label{acknowledgements}Acknowledgments}

Numerical computations were carried out on GPU cluster at the Center for Computational Astrophysics, National Astronomical Observatory of Japan. It is also partially supported by Recommendation Program for Young Researchers and Woman Researchers, Supercomputing Division, Information Technology Center, The University of Tokyo.
This work was supported by JSPS KAKENHI Grant Numbers 25K01052, 24K00688.
This research was supported by the grant of Joint Research by the National Institutes of Natural Sciences (NINS) (NINS program No OML032402).
This work was performed using the facilities of the Institute for Space–Earth Environmental Research (ISEE), Nagoya University.

\section*{Data Availability Statement}

The simulation codes that support the findings of this study are openly available in GitHub repository \url{https://github.com/keita-akutagawa}.
The simulation results that support the findings of this study are available from the corresponding author upon reasonable request.

\appendix

\section{Resolution dependence}

\begin{figure}[htbp]
    \includegraphics[width=0.95\linewidth]{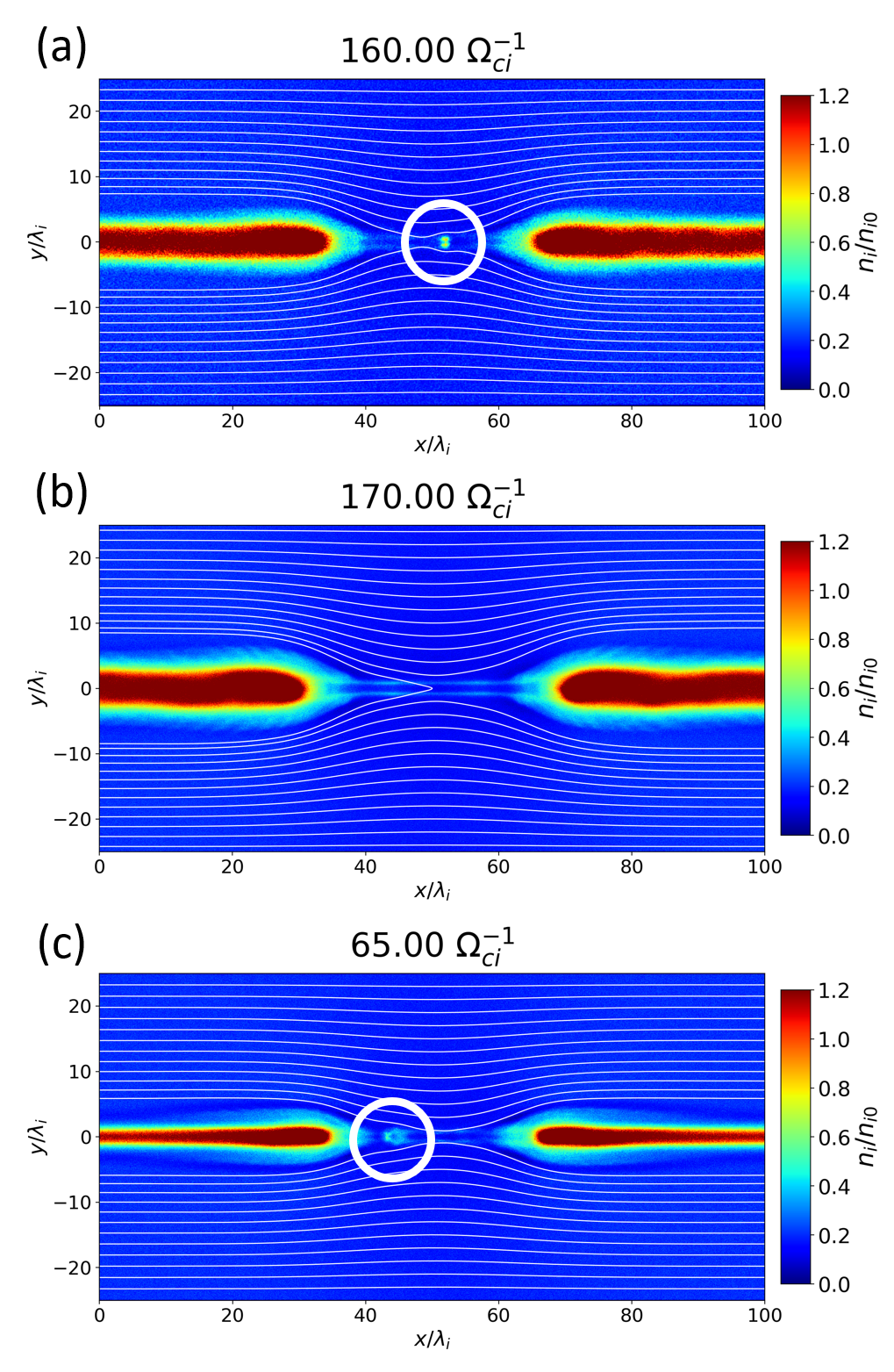}
    \caption{\label{appendix:resolution_dependence}Three simulation results of different $m_i / m_e$ and resolution. (a): $m_i / m_e = 16, \Delta = 0.5\lambda_e$, (b): $m_i / m_e = 16, \Delta = 0.1\lambda_e$, and (c): $m_i / m_e = 100, \Delta = 0.5\lambda_e$.}
\end{figure}

\rewrite{In this appendix, we examine the dependence of the PIC simulation results on grid resolution. Figure \ref{appendix:resolution_dependence} shows three cases with different $m_i / m_e$ and grid resolutions:
(a) $m_i / m_e = 16$, $\Delta = 0.5\lambda_e$;
(b) $m_i / m_e = 16$, $\Delta = 0.1\lambda_e$;
(c) $m_i / m_e = 100$, $\Delta = 0.5\lambda_e$. 
Here, $\Delta$ is the grid size. A comparison between panels (a) and (b) indicates that secondary plasmoids are formed when the electron inertial length $\lambda_e$ is resolved by only 2 grid points. On the other hand, there is no secondary plasmoid when $\lambda_e$ is resolved by 10 grids. A comparison between panels (b) and (c) indicates that secondary plasmoids also emerge for $m_i / m_e = 100$ case under the coarse resolution condition.
From these results, we conclude that sufficient grid resolution is essential for accurately capturing the formation of secondary plasmoids.}
\rewritesecond{Some previous studies report the formation of secondary plasmoids in PIC simulation. Our results in this appendix suggest the possibility that secondary plasmoids formed in previous studies may be due to the coarse resolution (under 10 grids for $1\lambda_e$). It is possible that secondary plasmoid could be formed under high $m_i / m_e$ due to different values of physical parameters ($\omega_{pe} / \Omega_{ce}, T_i / T_e$, and so on). It remains unclear and further simulations are needed.}

\bibliography{bib}

\end{document}